\documentclass[letterpaper,11pt]{article}
\usepackage[utf8]{inputenc}
\usepackage{amsmath}
\usepackage{amsthm}
\usepackage{mathrsfs}
\usepackage{fullpage}
\usepackage{color}
\usepackage{bbold}
\newtheorem{theorem}{Theorem}[section]
\newtheorem{corollary}{Corollary}[section]
\newtheorem{lemma}[theorem]{Lemma}
\newtheorem{fact}[theorem]{Fact}

\theoremstyle{definition}

\newtheorem{definition}{Definition}
\newtheorem{remark}{Remark}
\usepackage{algorithm}
\usepackage[noend]{algpseudocode}
\usepackage{graphicx}
\usepackage{url}
\usepackage{subcaption}
\definecolor{Blue}{rgb}{0.0,0.2,0.6}
\usepackage{hyperref}
\hypersetup{
    linktocpage=true,
    colorlinks=true,				
    linkcolor=Blue,				
    citecolor=Blue,				
    urlcolor=Blue,			
}

\newcommand{\argmin}{\mathop{\mathrm{arg\,min}}}
\newcommand{\dist}{\mathop{\mathrm{dist}}}

\newcommand{\F}{\mathop{\mathcal{F}}}

\newcommand{\ProjG}{\mathop{\mathrm{Proj}}\nolimits_{\Gamma}}

\algnewcommand{\IfThenElse}[3]{% \IfThenElse{<if>}{<then>}{<else>}
  \State \algorithmicif\ #1\ \algorithmicthen\ #2\ \algorithmicelse\ #3}
\title{Adaptive Privacy Budgeting}

\author{
Yuting Liang\thanks{Department of Computer Science, University of Toronto. \href{mailto:yliang@cs.toronto.edu}{yliang@cs.toronto.edu}}
\and
Ke Yi\thanks{Computer Science and Engineering, Hong Kong University of Science and Technology. \href{mailto:yike@cse.ust.hk}{yike@cse.ust.hk}}
}
\date{}
\begin{document}
\maketitle
\begin{abstract}
We study the problem of adaptive privacy budgeting under generalized differential privacy.
Consider the setting where each user $i\in [n]$ holds a tuple $x_i\in U:=U_1\times \dotsb \times U_T$, where $x_i(l)\in U_l$ represents the $l$-th component of their data. For every $l\in [T]$ (or a subset), an untrusted analyst wishes to compute some  $f_l(x_1(l),\dots,x_n(l))$, while respecting the privacy of each user. 
For many functions $f_l$, data from the users are not all equally important, and there is potential to use the privacy budgets of the users strategically, leading to privacy savings that can be used to improve the utility of later queries. In particular, the budgeting should be adaptive to the outputs of previous queries, so that greater savings can be achieved on more typical instances. In this paper, we provide such an adaptive budgeting framework, with various applications demonstrating its applicability.
\end{abstract}

\section{Introduction}
Differential Privacy (DP) \cite{dwork2006calibrating} is a popular model for protecting personal information that offers a rigorous privacy guarantee, and has been adopted and deployed by government agencies and large corporations \cite{appledp2017,erlingsson2014rappor,abowd2018us}. 
Roughly speaking, DP requires an algorithm to produce indistinguishable outputs on pairs of similar inputs, where similarity is defined in terms of the number of records that are different between the inputs. More generally, similarity between inputs can be measured in terms of a prespecified metric, where in this case the definition is also known as $d_{\chi}$-privacy \cite{chatzikokolakis2013broadening}, or Geo-Privacy \cite{liang2023concentrated} (GP).  Given a metric space $(U, \dist)$, GP requires the level of distinguishability between two inputs $x_i, x_i' \in U$ to be proportional to $\dist(x_i,x_i')$. The standard notion of DP corresponds precisely to choosing the Hamming metric as $\dist(\cdot,\cdot)$; however, this requirement is sometimes too strong, especially under the local model, where the input to the algorithm consists of just one individual's data. Under DP (or equivalently, GP using the Hamming metric), any two individuals' data must be indistinguishable, which requires injecting a large amount noise.  On the other hand, GP offers the flexibility to choose a metric appropriate to the data type and application.
Suppose the sensitive data consists of salary information, which depends mostly on the type and rank of the individual's job. In many cases, job types and ranks are public information, so adding a large amount of noise to mask the difference between salaries of two individuals with vastly different jobs might not be a good use of privacy; rather, stronger protection should be offered to individuals with similar jobs. Thus, choosing the Euclidean distance for salary data in the local model appears more suitable.

Both DP and GP enjoy a very important property known as \textit{composability}, which allows one to develop an algorithm by composing, possibly adaptively, multiple steps, while the privacy guarantee of the composed algorithm can be obtained by combining the privacy parameters of its constituent algorithms. 
The adaptivity of composability comes in two levels.  On the basic level, the choice of the next algorithm could depend on the outputs of all previous algorithms \cite{dwork2014algorithmic}, but the privacy parameters of all the algorithms must be fixed in advance.   However, having to fix privacy parameters in advance hinders optimal allocation of the privacy budget.  We would like a higher level of adaptivity, in particular, one that allows us to also adaptively set the privacy parameter for the next query; this is beneficial for analytical or monitoring tasks where depending on the outputs of the previous queries, the analyst may want a more or less accurate estimation for the next query.

Composability also allows us to track the privacy loss when the analyst wishes to ask multiple queries on the same set of users, each on a different attribute (e.g., location, financial, health data, etc), or ask the same query multiple times on a time-evolving attribute (e.g., location at different times).  To model both situations, or their mixture, with a unified framework, we consider the general setting where each user $i\in [n]$ holds a tuple $x_i\in U :=U_1\times \dotsb \times U_T$ for a possibly infinite $T$, where $x_i(l) \in U_l$ represents the $l$-th component of their data.  For each $l\in [T]$ and for each user $i\in [n]$, the untrusted analyst may apply a different DP/GP algorithm $M_{i,l}$ and receive as output $M_{1,l}(x_1(l)), \dots, M_{n,l}(x_n(l))$. 

With the power of composability described above, we are interested in the following question:  \textit{Given a privacy budget of $B$ for each user, how should an analyst allocate the budget for each user to maximize utility over a series of queries (each possibly about a different subset of the data components)?} 

Since the queries may change and adapt to the outputs on previous queries, we instead aim at answering the following question: \textit{Given a question of interest, and a pre-allocated privacy budget for it, how can one pose the query so that its privatization consumes the least amount of privacy, while maintaining a similar level of utility?}

In order to answer the question above, we develop a framework (in Section \ref{sec:privacy_filters} and \ref{sec:budgeting_frame}) that builds on several components, where each serves a different technical aspect of the problem:
\begin{enumerate}
    \item Generalized \textit{privacy filters} (Corollary~\ref{cor:fulladapt_comp_gpbound}, Theorems~\ref{thm:filter_cgp} and \ref{thm:approxgp_filter}), from the central model of DP \cite{rogers2016privacy} and CDP \cite{feldman2021individual}, to an arbitrary metric space and to many users; this allows one to choose both the GP algorithms and their privacy parameters adaptively, depending on the outputs of previous algorithms on all users.
    \item Consistent privacy accounting for heterogeneous data components (Theorem~\ref{thm:acc_compwise_gp}, Corollary~\ref{cor:acc_compwise_cgp}); this allows one to reason about budgeting across different data components.
    \item Template algorithms that allow \textit{privacy savings} (Algorithms~\ref{algo:pie_non}, \ref{algo:pie_topk}); privacy savings from earlier queries can be used for later queries when used together with a budgeting scheme (Algorithm~\ref{algo:mult_ie}) enabled by privacy filters.
\end{enumerate}
\begin{figure}[h]
     \centering
\includegraphics[width=0.48\textwidth]{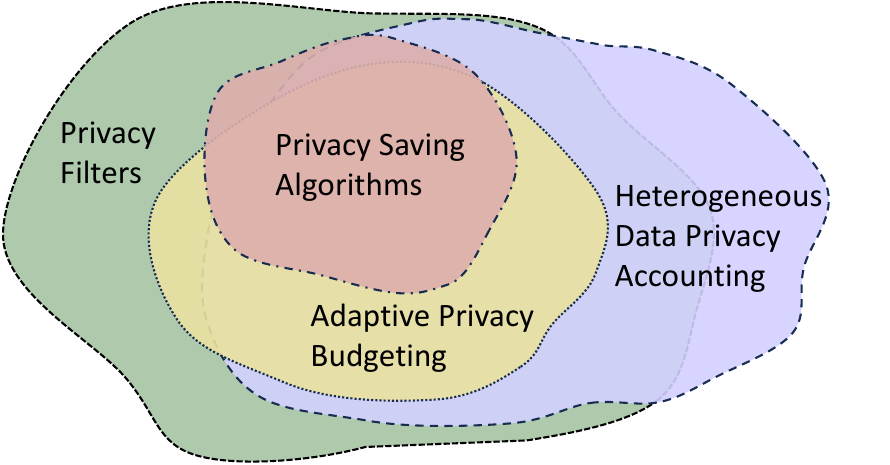}
            \label{fig:framework}
         \caption{Relationships among different components of our framework.
         }
\end{figure}
In Section~\ref{sec:adagp_app}, we show how our framework can be instantiated into privacy saving algorithms on several fundamental problems:  range counting, Gaussian kernel density estimation (KDE), and $k$ nearest neighbors ($k$NN) queries.   
For each of these applications, our framework allows more effective use of the privacy budget that adapts to the given instance, as demonstrated via experimental evaluation in Section~\ref{sec:experiments}.

Finally, we would like to point out that although our adaptive budgeting framework is developed from the perspective of the local model, it is equally applicable to the central model, where the data curator is considered a user holding some tuple $x_1$; in particular, it applies to the standard $(\varepsilon,\delta)$-DP setting, where the space $U$ is considered equipped with the Hamming metric. To avoid confusion, the applications discussed in the main text are considered under the local model. We discuss an application under the central model in Appendix~\ref{app:cdpcount}, along with experimental evaluation. Missing proofs to all theorems, lemmas and utility results also appear in the Appendix.

\section{Related Work}
\label{sec:related_work}
As mentioned, our framework generalizes the DP privacy filters \cite{rogers2016privacy,feldman2021individual}, in the sense that when setting $n=1$ and adopting the Hamming metric, our framework degenerates into them.  In addition to privacy filters, \cite{rogers2016privacy} has also proposed the notion of a \textit{privacy odometer}.  Roughly speaking, a privacy odometer is a function that computes a running upper bound on the total privacy loss of all the algorithms applied on the user.  It thus provides privacy guarantees that are \textit{ex-post}, i.e., the user can only be notified of their privacy loss after it has been incurred, which is weak.
In contrast, privacy filters, and also our budgeting framework, provide privacy guarantees that are \textit{ex-ante}, namely, for a specified privacy budget for each user, it is guaranteed that the total privacy loss will never exceed it, which is the same as the standard DP/GP definition.  

Our work is also related to the line of research in central-DP known as ``accuracy first'' mechanisms \cite{ligett2017accuracy,wu2019accuracy, whitehouse2022brownian}, which studies the problem of minimizing the privacy cost needed to satisfy a pre-specified level of accuracy. In these works, private algorithms are run until a certain level of accuracy is reached for some desired task, at which point a privacy cost can be calculated. Thus, these methods work under the privacy odometer framework and only provide ex-post guarantees. Another related line of works deals with heterogeneous privacy demands 
\cite{jorgensen2015conservative,yiwen2018utility,xue2022mean,sun2024personalized}, where the required privacy level is different for each user. In our framework, since users obtain different amounts of privacy savings for different queries, over time their remaining budgets vary; our algorithms also allow allocating heterogeneous budgets for a query.

There are several works related to the applications considered in this paper. 
The $k$NN query has been considered under GP in \cite{liang2023concentrated}, but in a different setting; there, their goal is to find, for a given user $i$ with a tuple of points $x_i$, $k$ points in $x_i$ that are closet to some point $p$. In our application, we consider identifying among all the the users $i\in [n]$, the $k$ of them with the closest distance to $p$.
The works in \cite{cormode2019answering,huang2021approximate,muthukrishnan2012optimal,wang2019answering} investigated range counting queries under DP, each belonging to a different class, while \cite{kaplan2020find, beimel2019private} studied how to find a point inside a convex hull. Range counting has also been considered under GP in \cite{liang2024smooth}, where they privatized one-way and two-way threshold functions using smooth sensitivity. Smooth sensitivity mechanisms aim to minimize the amount of noise needed for a query at each point, when they are allocated the same privacy budget; but they offer suboptimal solutions when many queries are desired, at the current state. It will be interesting to see whether and how our privacy budgeting framework can utilize smooth sensitivity mechanisms to further increase privacy savings. 

\section{Preliminaries}
\label{sec:prelim}
\subsection{Definitions, Notation and Mechanisms}
In differential privacy, we call a pair $x,x'\in U$ neighboring inputs if they differ by one individual's data, and write $x\sim x'$. DP is defined over all pairs of neighboring inputs:
\begin{definition} [Differential Privacy \cite{dwork2006calibrating}]
    Let $\varepsilon,\delta\ge 0$. A randomized mechanism $M:U\rightarrow V$ is $(\varepsilon,\delta)$-differentially private, or $(\varepsilon,\delta)$-DP, if for all measurable $S\subseteq V$ and all neighboring inputs $x,x'\in U$
    \[
\Pr[M(x)\in S] \le e^{\varepsilon}\Pr[M(x')\in S] + \delta.
\]
\end{definition}
In the special case where the input consists of data from only one individual, it is referred to as the local-DP model with abbreviation $(\varepsilon,\delta)$-LDP.

Let $(U,\dist)$ be a metric space. For $x,x'\in U$, we write $x\sim_{\Lambda}x'$ to denote $\dist(x,x')\le \Lambda$.
\begin{definition} [Geo-Privacy \cite{chatzikokolakis2013broadening,liang2023concentrated}]
\label{def:gp}
Let $\varepsilon, \delta \ge 0$ and $\Lambda\in\mathbb{R}_{>0}\cup\{\infty\}$. A randomized mechanism $M: U\rightarrow V$ is $(\varepsilon,\delta,\Lambda)$-geo-private, or $(\varepsilon,\delta,\Lambda)$-GP, if for all measurable $S\subseteq V$ and all $x\sim_{\Lambda}x'\in U$
\begin{equation}
\label{eqn:gp_def}
\Pr[M(x)\in S] \le e^{\varepsilon\cdot\dist(x,x')}\Pr[M(x')\in S] + \delta.
\end{equation}
\end{definition}
In the definition above, $\varepsilon$ measures privacy loss per unit distance (e.g. $0.001$ per meter for the Euclidean metric). We refer to the above as pure GP when $\delta=0$, and approximate GP otherwise.
We also write $(\varepsilon,0,\infty)$-GP as $\varepsilon$-GP for short. Note that $(\varepsilon,\delta,\Lambda)$-GP captures $(\varepsilon,\delta)$-DP as a special case: For central-DP, we use the Hamming distance $\dist_{\text{H}}(\cdot,\cdot)$, i.e. $(\varepsilon,\delta)$-DP is $(\varepsilon,\delta,1)$-GP with $\dist=\dist_{\text{H}}$; for local-DP, we use the discrete metric $\dist_{01}(\cdot,\cdot)$ defined by $\dist_{01}(x,x')=\mathbb{1}(x\neq x')$, i.e. $(\varepsilon,\delta)$-LDP is $(\varepsilon,\delta,1)$-GP with $\dist=\dist_{01}$. We set $\Lambda=1$ for DP since we only consider pairs $x,x'\in U$ with $\dist(x,x')\le 1$ as neighbors under DP. All remaining discussions will henceforth be made under the more general definition of GP, so all our results also hold for DP as a special case.

Another definition of geo-privacy is Concentrated Geo-Privacy (CGP), which enjoys better composability. It is defined in terms of R\'{e}nyi divergences:

\begin{definition} [R\'{e}nyi Divergences \cite{renyi1961measures, van2014renyi}] Fix $\alpha\in (0,1)\cup (1,\infty)$. For distributions $\mathcal{P}, \mathcal{Q}$ on space $V$ with pdf $p(\cdot), q(\cdot)$, respectively, the R\'{e}nyi divergence of order $\alpha$ is defined to be 
\begin{align*}
D_{\alpha}(\mathcal{P}\|\mathcal{Q}) &:=\frac{1}{\alpha-1}\log\left(\int_V p(z)^{\alpha}q(z)^{1-\alpha} dz\right) \\
&=\frac{1}{\alpha-1}\log\left(\mathbb{E}_{Z\sim \mathcal{P}}\left[\left(\frac{p(Z)}{q(Z)}\right)^{\alpha-1}\right]\right).
\end{align*}
The max-divergence is defined as 
\[
D_{\infty}(\mathcal{P}\|\mathcal{Q}):=\lim_{\alpha\rightarrow\infty} D_{\infty}(\mathcal{P}\|\mathcal{Q}) = \sup_{z\in V}\log\left(\frac{p(z)}{q(z)}\right).
\]
\end{definition}

\begin{definition} [Concentrated Geo-Privacy
\footnote{CGP was defined with $\Lambda=\infty$ in \cite{liang2023concentrated}, but the generalization to finite $\Lambda$ is straightforward.}
\cite{liang2023concentrated}]
\label{def:cgp}
Let $\rho\ge 0$ and $\Lambda\in\mathbb{R}_{>0}\cup\{\infty\}$. A randomized mechanism $M:U\rightarrow V$ is $(\rho,\Lambda)$-concentrated-geo-private, or $(\rho,\Lambda)$-CGP, if for all $x\sim_{\Lambda}x'\in U$ 
\begin{equation}
\label{eqn:cgp_def}
D_{\alpha}\left(\mathcal{M}(x)\|\mathcal{M}(x')\right) \le \rho\cdot \alpha\dist(x,x')^2,\;\; \forall \alpha > 1
\end{equation}
where $\mathcal{M}(x)$ and $\mathcal{M}(x')$ denote the distributions of the random variables $M(x)$ and $M(x')$, respectively.
\end{definition}
The parameter $\rho$ in CGP is the privacy loss per unit distance squared. We also write $(\rho,\infty)$-CGP as $\rho$-CGP for short.

Note that $M$ is $(\varepsilon,0,\Lambda)$-GP iff $D_{\infty}(\mathcal{M}(x)\|\mathcal{M}(x'))\le \varepsilon\dist(x,x')$ 
for all pairs of $x\sim_{\Lambda}x'\in U$. 
Moreover, these two versions of geo-privacy are related as follows:
\begin{lemma} [\cite{liang2023concentrated}]
\label{lm:gp_to_cgp}
A mechanism $M$ that is $(\varepsilon,0,\Lambda)$-GP is also $(\varepsilon^2/2,\Lambda)$-CGP.
\end{lemma}
\begin{lemma} [\cite{liang2023concentrated}]
\label{lm:cgp_to_approxgp}
    A mechanism $M$ that is $(\rho,\Lambda)$-CGP is also $(\varepsilon,\delta,\Lambda)$-GP, for all $\varepsilon,\delta, \Lambda$ satisfying
    \begin{equation}
    \label{eqn:approxgp}
    \varepsilon = \max\left\{\frac{s}{s-1}2\sqrt{\rho\log(2/(s+1)/\delta)},s\rho\Lambda\right\}
    \end{equation}
    for any $s>1$. In particular, choosing $s=1+\frac{2\sqrt{\rho\log(1/\delta)}}{\rho\Lambda}$ gives $(\rho\Lambda+2\sqrt{\rho\log(1/\delta)},\delta,\Lambda)$-GP.
\end{lemma}

We have the following composition results:
\begin{lemma}[\cite{andres2013geo,liang2023concentrated}]
Fix $r_1,\dotsb,r_k\ge 0$. For $j\in [k]$, let $M_j:U\rightarrow V_j$ be an algorithm with privacy parameter $r_j$. Let $M$ be a $k$-fold adaptive composition of the $M_j$'s. Then $M$ is
    \begin{enumerate}
        \item $(\sum_{j\in[k]}r_j)$-GP, if each $M_j$ is $r_j$-GP for $j\in [k]$;
        \item $(\sum_{j\in[k]}r_j)$-CGP, if each $M_j$ is $r_j$-CGP for $j\in [k]$.
    \end{enumerate}
\end{lemma}

For $f:U\rightarrow V\subseteq \mathbb{R}^d$, GP/CGP can be achieved as follows. Recall the function $f$ is $K$-Lipschitz if $\|f(x)-f(x')\|\le K \dist(x,x')$ for all $x,x'\in U$. 
\begin{lemma} [Basic Mechanisms \cite{chatzikokolakis2013broadening, liang2023concentrated}]
\label{lm:basic_mech}
    Let $f:U\rightarrow V\subseteq\mathbb{R}^d$ be a $K$-Lipschitz function. Let $Z=[Z_1,\dotsb,Z_d]^T$ be a random vector drawn from a distribution with pdf $\zeta(z)$.
    Then the mechanism $M(x):=f(x)+K\cdot Z$ is 
    \begin{itemize}
        \item[1.]$\varepsilon$-GP, if $\zeta(z)\propto e^{-{\varepsilon}\|z\|}$;
        \item[2.]$\rho$-CGP, if $\zeta(z)\propto e^{-{\rho}\|z\|^2}$, i.e. $Z\sim \mathcal{N}(0,\frac{1}{2\rho}I_{d\times d})$. 
    \end{itemize}
\end{lemma}

\section{Privacy Filters for GP}
\label{sec:privacy_filters}
In this section, we develop the privacy filter framework for geo-privacy, which allows us to develop GP algorithms with adaptive budgeting. 
We first introduce a few tools from stochastic processes, which are needed for and appropriately foreshadow the construction of the framework.

\subsection{Filtrations, Martingales and Stopping Times}
A collection of $\sigma$-algebras $\mathcal{F}:=\{\mathcal{F}_t\}_{t\ge 0}$ is called a \textit{filtration} if $\mathcal{F}_s\subseteq \mathcal{F}_l$ for all $s\le l$, so $\mathcal{F}_t$  models information available at time $t$. A stochastic process $Q$ is \textit{adapted} to the filtration $\mathcal{F}$ if $Q_t$ is $\mathcal{F}_t$-measurable (i.e., it can be determined by the information available at time $t$), and further \textit{predictable} w.r.t. $\mathcal{F}$ if it is $\mathcal{F}_{t-1}$-measurable, for all $t\ge 0$. 

A discrete-time process $Q$ is referred to as a \textit{martingale} (resp. \textit{supermartingale}) w.r.t. process $Y$ if for all $t\ge 0$ 
\begin{enumerate}
   \item $\mathbb{E}[|Q_t|]< \infty$; and  
   \item $\mathbb{E}[Q_{t+1}|Y_1,\dotsb,Y_t]= Q_t$ (resp. $\le Q_t)$.
\end{enumerate}

A stopping time $\tau$ is a random variable defined on a probability space $(\Omega,\mathcal{F},\mathbb{P})$, where $\tau:\Omega\rightarrow \mathbb{N}\cup \{\infty\}$ is such that $\{\tau = t\}\in \mathcal{F}_t$ for all $t\ge 0$, i.e., whether to stop at time $t$ can be decided by information available by time $t$. It is called a finite stopping time if $\tau < \infty$ almost surely. We define $\F_{\infty}:=\sigma(\cup_{t}\F_t)\subseteq \F$ when $\tau=\infty$. 
If $\tau_1$ and $\tau_2$ are both stopping times, then $\min(\tau_1,\tau_2)$ is also a stopping time.

For $t\ge 0$, denote $\tau \wedge t:= \min(\tau, t)$. We define the stopped process $Q^*:=\{Q_{\tau \wedge t}\}_{t\ge 0}$ where $Q^*_t=Q_t$ for $t\le \tau$, and $Q^*_t=Q_{\tau}$ for $t>\tau$. We will use the following result about stopped processes (e.g. see \cite{gallager2013stochastic,roch2024modern}):
\begin{theorem}
\label{thm:stopped_proc_supmart}
    Let $Q:=\{Q_t\}_{t\ge 0}$ be a supermartingale and $\tau$ be a stopping time, both w.r.t. a filtration $\mathcal{F}$. Then the stopped process $Q^*:=\{Q_{\tau \wedge t}\}_{t\ge 0}$ is also a supermartingale, and satisfies for all $t \ge 0$:
    \[
    \mathbb{E}[Q_t]\le \mathbb{E}[Q_{\tau \wedge t}] \le \mathbb{E}[Q_0].
    \]
\end{theorem}

\subsection{Filtering via Stopping Times}
\label{sec:filter_framework}
Consider the setting where $n$ users participate in a sequence of queries. At each step $t\ge 1$, we allow the analyst to select, possibly randomly, an algorithm $M_t$ (including its privacy parameters) and apply it on a subset of users $G_t$, based on the outputs of the previous $t-1$ queries.  Users not in $G_t$ can be considered as being applied with the constant algorithm $M_{\mathrm{NULL}}(\cdot)\equiv\mathrm{NULL}$.

Let $Y_{[n],t}:=(Y_{1,t},\dotsb,Y_{n,t})$ denote the random vector of outputs of the $t$-th query from the $n$ users for $t\ge 1$. We further write $Y_{[n],[t]}:=(Y_{[n],1},\dotsb,Y_{[n],t})$ for the collection of outputs from all $t$ queries. The information available to the analyst at time $t$ is the $\sigma$-algebra generated from $Y_{[n],[t]}$, plus any coin tosses $W_{[t]}$ the analyst uses for selecting the sequence of algorithms up to time $t$, i.e., $\mathcal{F}_t=\sigma(Y_{[n],[t]}, W_{[t]})$ for $t\ge 1$. At time $t-1$, $M_t=M_t(w_{t-1},y_{[n],[t-1]})$ is selected based on the coin toss $w_{t-1}$ and all previous outputs $y_{[n],[t-1]}$.  Define $\mathcal{F}_0:=\sigma(W_0)$.
Then, the following variables are all $\F_{t-1}$-measurable: 
\begin{enumerate}
    \item[1)] the outcomes of $M_{t-1}$ applied to $x_1,\dotsb, x_n$;
    \item[2)] the \textit{choice} of algorithm $M_t$; 
    \item[3)] the worst-case privacy consumption $M_t$; and 
    \item[4)] the subset of users $G_t$ that receive $M_t$. 
\end{enumerate}
For $z_t$ in the range of $M_t$, denote the pdf of $M_t$ when applied to $x$ as $m_t(x;w_{t-1},y_{[n],[t-1]})(z_t)$ . Moreover,  we can write the joint pdf for the composed sequence $M_{[t]}=(M_1,\dotsb,M_t)$ applied to $x$ at $z_{[t]}:=(z_1,z_2,\dotsb,z_t)$ as
\[
m_{[t]}(x;w_{[t-1]},y_{[n],[t-1]})(z_{[t]})=m_1(x;w_0)(z_1)\dotsb m_t(x;w_{t-1},y_{[n],[t-1]})(z_t)\]
where for all $i\in [n]$, $y_{i,1}:=M_1(w_0)(x_i)$, $\dots$, $y_{i,t-1}:=M_{t-1}(w_{t-2},y_{[n],[t-2]})(x_i)$.

We introduce two functions for privacy accounting (i.e. tracking privacy loss).  
Define a function $\mathcal{E}_{\Lambda}$ which takes as input an algorithm $M$, and computes
\begin{equation}
\label{eqn:gp_worstcase_acc}
\mathcal{E}_{\Lambda}(M):=\sup_{x\sim_{\Lambda}x'} \frac{D_{\infty}(\mathcal{M}(x)\|\mathcal{M}(x'))}{\dist(x,x')}.
\end{equation}
Similarly, we define a function $\mathcal{R}_{\Lambda}$ for CGP:
\begin{equation}
\label{eqn:cgp_worstcase_acc}
\mathcal{R}_{\Lambda}(M):=\sup_{x\sim_{\Lambda}x'} \sup_{\alpha > 1} \frac{D_{\alpha}\left(\mathcal{M}(x)\|\mathcal{M}(x')\right)}{\alpha \dist(x,x')^2}.
\end{equation}
Then every $M$ satisfies $(\mathcal{E}_{\Lambda}(M),0,\Lambda)$-GP and $(\mathcal{R}_{\Lambda}(M),\Lambda)$-CGP. 
We adopt the convention that ${0}/{0}=0$. Thus, for the constant algorithm $M_{\mathrm{NULL}}(\cdot)$, which is perfectly private, we have $\mathcal{E}_{\Lambda}(M_{\mathrm{NULL}})=\mathcal{R}_{\Lambda}(M_{\mathrm{NULL}})=0$ for all $\Lambda$. 

For an algorithm $M:U\rightarrow V$, the two functions above are related via a function $L:V\rightarrow \mathbb{R}$:
% To capture the privacy loss of an algorithm $M:U\rightarrow V$ at a particular output, we introduce a function $L:V\rightarrow \mathbb{R}$.  
For any $x,x'\in U$ and $z\in V$, define $L$ as
\begin{equation}
\label{eqn:log_loss}
L(M,x,x')(z):=\log\left(\frac{m(x)(z)}{m(x')(z)}\right),
\end{equation}
where $m(x)(\cdot)$ and $m(x')(\cdot)$ are the pdf's of the distributions $\mathcal{M}(x)$ and $\mathcal{M}(x')$, respectively.  Then, \eqref{eqn:gp_worstcase_acc} and \eqref{eqn:cgp_worstcase_acc} can be equivalently re-stated as:
% Note that the worst-case privacy consumptions of $M$ relate to $L$ as follows:
\begin{align*}
\mathcal{E}_{\Lambda}(M)&=\sup_{x\sim_{\Lambda}x'} \left(\frac{\sup_{z\in V} L(M,x,x')(z)}{\dist(x,x')}\right), \\
\mathcal{R}_{\Lambda}(M)&=\sup_{x\sim_{\Lambda}x'} \left(\sup_{\alpha>1}\frac{ \log\left(\mathbb{E}_{Z\sim \mathcal{M}(x)}\left[e^{(\alpha-1)L(M,x,x')(Z)}\right]\right)}{(\alpha-1)\cdot\alpha\dist(x,x')^2}\right).
\end{align*}

\subsubsection{Filters for pure GP}

We first note the following \textit{fully adaptive} composition result for GP.

\begin{theorem}
\label{thm:fulladapt_comp_GP}
Fix $t\in \mathbb{N}$ and $\Lambda\in \mathbb{R}_{>0}\cup \{\infty\}$. Let $M_{[t]}$ denote a sequence of potentially adaptive algorithms. Fix a path $(w_{[t-1]},y_{[n],[t-1]})$
where $y_{i,j}$ is in the range of ${M}_j(w_{j-1},y_{[n],[j-1]})$ for $i\in [n]$.
Then for every $z_{[t]}=(z_1,\dotsb,z_t)\in V_1\times \dotsb \times V_t$, and every pair of $x\sim_{\Lambda}x'\in U$ it holds that
\[L(M_{[t]},x,x')(z_{[t]})\le \dist(x,x')\sum_{j=1}^t \mathcal{E}_{\Lambda}({M}_j(w_{j-1},y_{[n],[j-1]})),\]
where each $M_j=M_j(w_{j-1},y_{[n],[j-1]})$ for $j\in [t]$ and $M_{[t]}=(M_1,\dotsb,M_t)$ .
\end{theorem}
The proof is straightforward and in Appendix~\ref{appendix:fulladapt_comp_GP}.

\begin{corollary}
\label{cor:fulladapt_comp_gpbound}
The composition $M_{[t]}$ of the sequence of adaptive algorithms is $(\epsilon_\Lambda(W_{[t-1]}, Y_{[n],[t-1]}), 0, \Lambda)$-GP, where
\[
\epsilon_\Lambda(W_{[t-1]}, Y_{[n],[t-1]}):=\sum_{j=1}^t \mathcal{E}_{\Lambda}\left(M_{j}(W_{j-1},Y_{[n],j-1})\right).
\]
\end{corollary}

The corollary above gives us a bound on the privacy parameter of $M_{[t]}$ based on the random vector of coin tosses $W_{[t-1]}$ and collection of random vectors $Y_{[n],[t-1]}$ corresponding to the algorithm outputs from all the users 
up to time $t$. Thus, the resulting privacy guarantee is probabilistic or \textit{ex-post}. In order to have a deterministic guarantee that works for all possible paths $(w_{[t-1]},y_{[n],[t-1]})$, or is \textit{ex-ante}, we introduce an object known as a \textit{privacy filter} which constrains the possible length and sequence of algorithms that we can apply to a user's data.

A privacy filter is a stopping rule which ensures that the privacy consumption of a sequence of algorithms does not go over a pre-specified budget $B$. Fix a user $i$ and a path $(w_{[t-1]},y_{[n],[t-1]})$. Let $J(t)$ denote the indices $j$ such that $M_j$ has been applied to $i$ up to and including time $t$.
\begin{algorithm}[h]
        \caption{Local protocol $\mathrm{P}$ with $F_B$: local view at $x$}
        \label{algo:local_view_randomtau}
            \begin{flushleft}
            \textbf{Input}: $M_l$ at time $l$\\
            \textbf{Output}: $(b_l, y_l)$ at time $l$
            \end{flushleft}
            \begin{algorithmic}[1]
             \State initialize $b\gets \mathrm{CONT}$
            \If{receive $M_l$ at time $l$}
            \If{$b=\mathrm{HALT}$} 
            \State {$y_{l} \gets \mathrm{NULL}$,\; $r_{l} \gets 0$}
             \Else 
             \State {$R_l \gets \mathcal{E}_\Lambda(M_l)$ or $\mathcal{R}_\Lambda(M_l)$ }%
             \If{$R_l=0$}
             \State $y_{l} \gets \mathrm{NULL}$,\; $r_{l} \gets 0$
                \ElsIf{$F_B(r_{1},\dotsb,r_{l-1},{R}_{l})= \text{CONT}$}
                \State $r_{l} \gets {R}_{l}$
                \State compute $y_{l} \gets M_{l}(x)$
                \Else
                \State $b\gets \mathrm{HALT}$ 
                \State $y_{l} \gets \mathrm{NULL}$,\; $r_{l} \gets 0$
                \EndIf
            \EndIf
             \State output $(b, y_{l})$
            \EndIf
            \end{algorithmic}
        \end{algorithm}

For each time $j\in [t]$, if the algorithm $M_j$ has not been applied to $i$, we assume user $i$ has received $M_{\mathrm{NULL}}$ at time $j$,
where the pdf of $M_{\mathrm{NULL}}$ is $1$ for the value $\mathrm{NULL}$ and $0$ otherwise. 
Thus, the joint pdf $m_{[t]}(x)$ at any $z_{[t]}$ is equal to the product of pdf's arising from each access to $x$ by applying $M_j$ for $j\in J(t)$, or zero.
If $m_{[t]}(x)(z_{[t]})\neq 0$, then $m_{[t]}(x)(z_{[t]})=m_{J(t)}(x)(z_{J(t)})$, where
\[
m_{J(t)}(x)(z_{J(t)}) := \Pi_{j\in J(t)} m_j(x;w_{j-1},y_{[n],[j-1]}).
\]
Then, by Theorem~\ref{thm:fulladapt_comp_GP}
\begin{align*}
&{}\log\left(\frac{m_{[t]}(x;w_{[t-1]},y_{[n],[t-1]})(z_{[t]})}{m_{[t]}(x';w_{[t-1]},y_{[n],[t-1]})(z_{[t]})}\right) \le \dist(x,x')\cdot\sum\nolimits_{j\in J(t)} \mathcal{E}_{\Lambda}\left(M_j(w_{j-1},y_{[n],[j-1]}\right).
\end{align*}
The RHS of the above inequality is bounded by $B\cdot \dist(x,x')$ if $\sum_{j\in J(t)} \mathcal{E}_{\Lambda}\left(M_j(w_{j-1},y_{[n],[j-1]}\right)\le B$, and this holds for any user $i$, any pair $x,x'$ and any arbitrary path. Thus, we can define a simple stopping rule using a function  $F_{B,\infty}:[0,B]^* \rightarrow \{\text{CONT},\mathrm{HALT}\}$ for a pre-specified budget $B$, where for all $t\in\mathbb{N}$:
\begin{equation}
\label{eqn:F_B_0_inf}
F_{B,\infty}(r_1,\dotsb,r_t) = \begin{cases}
\text{CONT},  \text{\;\;\;if\;} \sum_{j=1}^t r_j \le B\\
\mathrm{HALT}, \text{\;\;\;else.}
\end{cases}
\end{equation} 
A local protocol which implements a privacy filter is given in Algorithm~\ref{algo:local_view_randomtau}. The stopping rule defined above ensures that $\sum_{j\in J(t)}r_j\le B$ for all $t\in\mathbb{N}$. Thus, 
Algorithm~\ref{algo:local_view_randomtau} with the filter $F_{B,\infty}$ defined in equation \eqref{eqn:F_B_0_inf} satisfies $(B,0,\Lambda)$-GP, provided that the sequence of algorithms are determined by the interactions between the analyst and the users as outlined in Algorithm~\ref{algo:template}, where the privacy parameters $R_j$'s are computed as:
\begin{equation}
\label{eqn:privacy_step_bound}
R_j(w_{j-1}, y_{[n],[j-1]}):=\mathcal{E}_{\Lambda}({M}_j(w_{j-1},y_{[n],[j-1]})).
\end{equation}
\begin{algorithm}[h]
    \caption{Template Algorithm for Local Privacy Filter}
    \label{algo:template}
        \begin{flushleft}
        \textbf{Input}: $(x_1,\dotsb,x_n)\in U^n$; initial budget $B$ for all users\\
        \textbf{Output}: variable-length sequence $y_{[n],1} y_{[n],2}, \dotsb$
        \end{flushleft}
        \begin{algorithmic}[1]
        \State $B_i \gets B$ for all $i\in [n]$ \Comment{initial privacy budget for all $i$}
        \For{$j=1,\dotsb$}
            \State $w_{j-1} \gets$ coin tosses drawn by Analyst
            \State Analyst selects $M_{j}$, privacy parameter $R_{j}$, and subset $G_j$%
            \For{$i\in G_{j}$}%
            \State Analyst sends $M_{j}$ to user $i$, then receives $(b_i, y_{i,j})$ 
            \If{$b_i = \text{CONT}$}
            \State $B_i\gets B_i - R_{j}$ \Comment{updated (worst-case) budget for $i$}%
            \EndIf
            \EndFor
            \State Output $y_{[n],j}:=(y_{1,j},\dotsb,y_{n,j})$
        \EndFor
        \end{algorithmic}
    \end{algorithm}

Algorithm~\ref{algo:template} gives the ``global'' view of the analyst. Note that in Algorithm~\ref{algo:template}, the analyst can in fact select a different algorithm with different (worst-case) privacy consumption for each $i$; this is equivalent to selecting just index $i$ in $G_j$. 

\begin{remark} Although Algorithm~\ref{algo:template} specifies the set of users eligible for participation as an input, in practice we do not need to have them specified at the start of algorithm since users can have their budgets initialized at the first time they are selected to be queried. The interaction would stop when all the queried users have exhausted their budgets and the analyst does not have a query for new users.
\end{remark}

\begin{remark}  In line 13 of Algorithm~\ref{algo:local_view_randomtau}, $b$ is changed to $\mathrm{HALT}$ the first time the privacy filter returns $\mathrm{HALT}$, this makes all subsequent outputs equal to $(\mathrm{HALT},\mathrm{NULL})$; in fact, this is not necessary by the adaptive composition result of Corollary~\ref{cor:fulladapt_comp_gpbound}, and it's introduced to only make the analysis in the next subsection cleaner. Nevertheless, the analyst has control over when a user drops out from future queries since he/she can compute the remaining privacy budget of any user at any time.
\end{remark}

\subsubsection{Filters for CGP}
The privacy filter result for GP algorithms presented in the previous subsection is a simple consequence of the fully adaptive composition result from Corollary~\ref{cor:fulladapt_comp_gpbound}; to prove the latter, 
it suffices to bound $L(M,x,x')(z)$ pointwise for each $z$ in the range of $M$ for arbitrary pairs $x\sim_{\Lambda} x'$, 
which yields a straightforward calculation that extends easily to compositions $M_{[t]}=(M_1,\dotsb,M_t)$ for arbitrary $t\ge 1$ or any countable sequence of $M_j$'s using an inductive argument. 

For CGP, the situation is more subtle as we need to bound $\mathbb{E}_{Z\sim \mathcal{M}(x)} [e^{(\alpha-1)L(M,x,x')(Z)}]$.
Since the privacy filter gives an ex-ante guarantee, this means the expectation is taken w.r.t. to $\mathcal{F}_0$. {Fortunately, we can use Theorem~\ref{thm:stopped_proc_supmart} to help calculate the required expectations, where the stopping time $\tau$ is the time when access to the data at the local view (defined in Algorithm~\ref{algo:local_view_randomtau}) is cut off. We first state a general result on the composition of the sequence of $M_j$'s.}
\begin{theorem}
\label{thm:cgp_comp_bound}
Fix $B\ge 0$, $\Lambda\in \mathbb{R}_{>0}\cup \{\infty\}$ and $t\in\mathbb{N}$. 
Let $M_{[t]}$ be a potentially adaptive sequence of algorithms selected by the analyst as in Algorithm~\ref{algo:template}.
Suppose the $M_j$'s are selected such that $\sum_{j=1}^t R_j \le B$ for some $B\ge 0$,
where each $R_j$ is computed as
\begin{align}
\label{eqn:privcgp_step_bound}
R_j(w_{j-1}, y_{[n],[j-1]}) &= \mathcal{R}_{\Lambda}\left(M_j(w_{j-1}, y_{[n],[j-1]})\right).
\end{align}
Then $M_{[t]}$ satisfies $(B,\Lambda)$-CGP.
\end{theorem}

{
Next, we describe the construction of $\tau$. 
The interactions in Algorithm~\ref{algo:template} can continue for as long as the analyst wishes, which can be a predetermined number of rounds, or the analyst can decide to stop following a rule that depends on the coin tosses and the outputs from all the users of all the queries asked up to the current time. Thus, the time $\tau(A)$ when the analyst stops making queries is a stopping time. At the local view defined in Algorithm~\ref{algo:local_view_randomtau}, the time when the privacy filter function $F_B$ first outputs $\mathrm{HALT}$ is also a stopping time. Thus, $\tau$ is the minimum of these two times.

It turns out that the privacy filter $F_{B,\infty}$ defined in equation~\eqref{eqn:F_B_0_inf} for GP algorithms also works for CGP.}
\begin{theorem}
\label{thm:filter_cgp}
Fix $B\ge 0$, $\Lambda\in \mathbb{R}_{>0}\cup \{\infty\}$.
Then Algorithm~\ref{algo:local_view_randomtau} implemented with the filter $F_{B,\infty}$ defined in equation~\eqref{eqn:F_B_0_inf} satisfies $(B,\Lambda)$-CGP when used as the local protocol to Algorithm~\ref{algo:template}, where $R_j$ is computed as in equation~\eqref{eqn:privcgp_step_bound}.
\end{theorem}

{
\begin{remark}
    Note that previously under the central-DP setting, a large number $N\in \mathbb{N}$ is set at the beginning as the maximum number of queries the analyst can ask \cite{feldman2021individual}; this can be achieved by choosing $\tau(A):=N$ in our framework, but here we allow it to be random.
\end{remark}
}

\subsubsection{Filters for approximate GP}
We have by Lemma~\ref{lm:cgp_to_approxgp} that $(\rho,\Lambda)$-CGP implies $(\varepsilon,\delta,\Lambda)$-GP where $\varepsilon$ is given by equation \eqref{eqn:approxgp}. Thus, from Theorem~\ref{thm:filter_cgp} the following is immediate:
\begin{corollary}
    Algorithm~\ref{algo:local_view_randomtau} implemented with the filter $F_{B.\infty}$ defined in equation~\eqref{eqn:F_B_0_inf} satisfies $(B',\delta,\Lambda)$-GP when used as the local protocol to Algorithm~\ref{algo:template}, with $R_j$ as given in equation \eqref{eqn:privcgp_step_bound} and $B'=B\Lambda+2\sqrt{\rho\log(1/\delta)}$.
\end{corollary}
Alternatively, we can use the following filter which provides tighter accounting:
    Fix $B\ge 0$, $\delta,\Lambda>0$. Let $g_{\delta}:s\mapsto \frac{s}{s-1} 2 \sqrt{\log\frac{2}{(s+1)\delta}}$. 
    Let $F^{\delta,\Lambda}_{B,\infty}$ be 
    defined on all sequences $r\ge 0$ of length $t$ for all $t\in \mathbb{N}$:
    \begin{align}
        \label{eqn:approxgp_filter}
&{}F^{\delta,\Lambda}_{B,N}(r) 
= \begin{cases}
                \mathrm{CONT},  \mathrm{\;if\;}\underset{s>1}{\min} \max \left(g_{\delta}(s)\sqrt{{\sum}_{j\in [t]}r_j}, s\Lambda {\sum}_{j\in [t]} r_j\right) \le B \\
                \mathrm{HALT}, \mathrm{\;else}.
            \end{cases}
    \end{align}

\begin{theorem}
\label{thm:approxgp_filter}
    Algorithm~\ref{algo:local_view_randomtau} implemented with filter $F^{\delta,\Lambda}_{B,\infty}$ defined in equation~\eqref{eqn:approxgp_filter} satisfies $(B,\delta,\Lambda)$-GP when used as the local protocol to Algorithm~\ref{algo:template}, where $R_j$ is computed as 
    in equation~\eqref{eqn:privcgp_step_bound}.
\end{theorem}

\section{The Privacy Budgeting Framework}
\label{sec:budgeting_frame}
The privacy filter framework developed in the previous section can be applied to any arbitrary data domain equipped with a metric. 
Suppose our domain is $U:=U_1 \times U_2 \times \dotsb \times U_T$ where each $U_l$ is a metric space equipped with metric $\dist_l$, for $l\in [T]$. We can define a metric $\dist_{\infty}$ on $U$ where $\dist_{\infty}:U\rightarrow \mathbb{R}_{\ge 0}$ is defined by $\dist_{\infty}(x,x')=\max_{l\in [T]} \dist_l(x(l),x'(l))$. For a tuple $x$ in a product space $U$, the $\dist_{\infty}$ metric offers the strongest privacy guarantee, since it provides protection for each component in $x$. Moreover, when all the queries are made to the same component, the case reduces to the classic setting with a single data component. 
The $\dist_{\infty}$ metric has been adopted by \cite{andres2013geo,liang2023concentrated} for the special case where $U_l=U_1\subseteq \mathbb{R}^d$ for $l\in [T]$. In particular, if the $l$'s correspond to time steps, then $U$ contains time series data. 

Here, we allow $U$ to also contain heterogeneous data, where data in different components can reside in different spaces. For example, the first component could contain location information, while the second could contain health or financial data, etc. In this case, if we apply an $\varepsilon_1$-GP mechanism $M_1:U_1\rightarrow V_1$ (w.r.t. $\dist_1$) to the first component and another $\varepsilon_2$-GP mechanism $M_2:U_2\rightarrow V_2$ (w.r.t. $\dist_2$) to the second component, the composition  $(M_1,M_2)$ satisfies $(\varepsilon_1+\varepsilon_2)$-GP on $U$ w.r.t. $\dist_{\infty}$, since $\varepsilon_1\dist_1(x,x')+\varepsilon_2\dist_2(x,x')\le (\varepsilon_1+\varepsilon_2)\cdot\max\left(\dist_1(x,x'),\dist_2(x,x')\right) \le (\varepsilon_1+\varepsilon_2)\dist_{\infty}(x,x')$ for any pair $x, x'$. More specifically, privacy accounting for $U$ equipped with $\dist_{\infty}$ works in the most intuitive way:
\begin{theorem}
\label{thm:acc_compwise_gp}
Fix $l\in [T]$. Let $M:U\rightarrow V$ be an algorithm that depends only on the $l$-th component, i.e., $M(x)=M(x')$ for all $x, x'\in U$ with $x(l)=x'(l)$. Then
\begin{align*}
\mathcal{E}_{\Lambda}(M) &:= \sup_{u\sim_{\Lambda} u'\in U} \frac{D_{\infty}\left(\mathcal{M}(u)\|\mathcal{M}(u')\right)}{\dist_{\infty}(u,u')}\\
&= \sup_{z\sim_{\Lambda} z'\in U_l} \frac{D_{\infty}\left(\mathcal{M}^{(l)}(z)\|\mathcal{M}^{(l)}(z')\right)}{\dist_{l}(z,z')}
=: \mathcal{E}_{\Lambda}(M^{(l)})
\end{align*}
where $M^{(l)}:U_l\rightarrow V$ is defined by $M^{(l)}(z)=M\left(x\right)$ for any $x\in U$ with $x(l)=z\in U_l$.
\end{theorem}
 \begin{proof}
     $D_{\infty}\left(\mathcal{M}(u)\|\mathcal{M}(u')\right)=D_{\infty}\left(\mathcal{M}^{(l)}(u(l))\|\mathcal{M}^{(l)}(u'(l))\right)$ and $\dist_{l}(u(l),u'(l))\le \dist_{\infty}(u,u')$, so 
     \begin{align*}
     \sup_{u\sim_{\Lambda} u'\in U} \frac{D_{\infty}\left(\mathcal{M}(u)\|\mathcal{M}(u')\right)}{\dist_{\infty}(u,u')} 
     &\le  \sup_{u\sim_{\Lambda} u'\in U} \frac{D_{\infty}\left(\mathcal{M}^{(l)}(u(l))\|\mathcal{M}^{(l)}(u'(l))\right)}{\dist_{l}(u(l),u'(l))}\\
     &\le \sup_{z\sim_{\Lambda} z'\in U_l} \frac{D_{\infty}\left(\mathcal{M}^{(l)}(z)\|\mathcal{M}^{(l)}(z')\right)}{\dist_{l}(z,z')}.
     \end{align*}
For the other direction, fix any $z\sim_{\Lambda} z'\in U_l$. For any arbitrary $z_j\in U_j$ for $j\in [T]\setminus\{l\}$, let $x=(z_1,\dotsb,z_{l-1},z,z_{l-1},\dotsb,z_T)$, and $x'$ be obtained by replacing the $l$-th component of $x$ with $z'$. Then $x,x'\in U$ and $\dist_{\infty}(x,x')=\dist_{l}(x(l),x'(l))=\dist_l(z,z')\le \Lambda$, so
\begin{align*}
    \frac{D_{\infty}\left(\mathcal{M}^{(l)}(z)\|\mathcal{M}^{(l)}(z')\right)}{\dist_{l}(z,z')} &= \frac{D_{\infty}\left(\mathcal{M}(x)\|\mathcal{M}(x')\right)}{\dist_{\infty}(x,x')}\\
    &\le \sup_{u\sim_{\Lambda} u'} \frac{D_{\infty}\left(\mathcal{M}(u)\|\mathcal{M}(u')\right)}{\dist_{\infty}(u,u')}.
\end{align*}
Since the above holds for any arbitrary pair $z\sim_{\Lambda} z'\in U_l$, taking $\sup$ over $z\sim_{\Lambda} z'\in U_l$ on the left hand side gives the desired result.
 \end{proof}
Thus, privacy accounting w.r.t. $\dist_{\infty}$ for queries that depend on a single component reduces to accounting component-wise w.r.t. to its associated metric. By similar arguments, we also have:
\begin{corollary}
\label{cor:acc_compwise_cgp}
Fix $l\in [T]$. Let $M:U\rightarrow V$ be an algorithm that depends only on the $l$-th component, i.e., $M(x)=M(x')$ for all $x, x'\in U$ where $x(l)=x'(l)$. Then
\[
\mathcal{R}_{\Lambda}(M) = \mathcal{R}_{\Lambda}(M^{(l)})
\]
where $M^{(l)}:U_l\rightarrow V$ is defined by $M^{(l)}(z)=M\left(x\right)$ for any $x\in U$ with $x(l)=z\in U_l$.
\end{corollary}

\begin{remark}
    Note that by Theorem~\ref{thm:acc_compwise_gp} and Corollary~\ref{cor:acc_compwise_cgp}, we can in fact have $U:=\Pi_{l=1}^{\infty} U_l$ and the components need not be specified in advance; the pair $(U_l,\dist_l)$ needs to be specified only when the component $U_l$ is to be queried for the first time, for privacy accounting, and our GP/CGP filters permit querying users' data indefinitely for as long as their budgets allow. Moreover, for data that possibly change over time, the changed data can simply be treated as new components when the analyst wants to query them.
\end{remark}
\subsection{Budgeting for Multiple Queries and Users}
\label{sec:mult_q}
Suppose we want to make $m$ sequential queries to components at indices ${\theta_1,\dotsb,\theta_m}\subseteq [T]$ (e.g. for time series data each component corresponds to a different time step), and each user has an initial privacy budget of $B_i=B$. Our budgeting framework aims to answer the following question: {How can we allocate the budget $B_i$ to the $m$ queries for each user $i$, so that the overall utility is maximized, when the remaining queries are potentially unknown until the time they are made and adaptive to the outputs of the previous queries?}

Without adaptive budgeting, the analyst can use for each user $i\in [n]$ a uniform budget of $r_i=B/m$ to privatize each of the $m$ queries, or apply a fraction $\omega_l$ of $B$ to the $l$-th query, for some combination of $\omega_l$'s with $\omega_1+\dotsb+\omega_m=1$. In any case, without adaptive budgeting, the budget allocation must be fixed at the beginning. 

When we implement the queries as part of a privacy filter framework, we can adaptively budget for the next query; 
our idea is to try to minimize the privacy consumption of the each query. 
Suppose for the first query we had initially allocated a budget of $r_i=B/m$ to every user $i$, and we have an algorithm that computes the query with the same utility while allowing a subset of users to incur a privacy cost less than $B/m$. After completing the first query, every user $i$ has a remaining budget of $B_i\ge B-B/m$, i.e., incurs a budget saving of $B_i-(m-1)B/m\ge 0$.
One can utilize the privacy savings in different ways; for example, we can evenly distribute it among the remaining $m-1$ queries, i.e., re-allocate a maximum budget of $B_i/(m-1)$ for the next query for user $i$. This method favors later queries since they accumulate more savings than the earlier queries. Alternatively, we can utilize a proportion (e.g. $3/4$) of the savings immediately, allowing the next query to benefit more from the privacy savings incurred so far. If $m$ is fixed, we can also increase the proportion linearly from $3/4$ for the second query to $1$ for the last query. 
We give an example implementation of this budgeting process in Algorithm~\ref{algo:mult_ie}.
$\Psi_l$ contains the parameters for the $l$-th query, and $\mathrm{PSA}_l$ is an algorithm for the $l$-th query that allows privacy savings.
\begin{algorithm}
    \caption{Multiple Queries utilizing Privacy Savings}
    \label{algo:mult_ie}
    \begin{flushleft}
        \textbf{Input}: $\Psi_1,\dotsb,\Psi_m$; $x_{1},\dotsb,x_{n}$; privacy budget $B>0$ \\
        \textbf{Output}: outputs of the $m$ queries $a_1, a_2, \dotsb, a_m$ 
         \end{flushleft}
        \begin{algorithmic}[1]
        \State $B_i \gets B$ for $i\in [n]$
        \For{$l=1,2,\dotsb, m$}
            \State $\gamma_l \gets \frac{(l-1)}{(m-1)}(1-3/4)+3/4$
            \State $r_i \gets \gamma_l\left(B_i-\frac{m-(l-1)}{m}B\right)+\frac{B}{m}$ for $i\in [n]$
            \State $(a_l, B_{[n]}) \gets  \mathrm{PSA}_l(\Psi_s)((x_{1}(\theta_l),\dotsb,x_{n}(\theta_l),r_{[n]},B_{[n]})$ 
        \EndFor
        \State output $a_1,\dotsb,a_m$
        \end{algorithmic}
    \end{algorithm}

\subsection{Privacy Saving via Iterative Elimination}
In this subsection, we develop algorithms which allow privacy savings that can be used in the budgeting process described in Algorithm~\ref{algo:mult_ie}. I.e., they are to be used as $\mathrm{PSA}_l$ in Algorithm~\ref{algo:mult_ie}, where a query is made to $\theta_l$-th component.

For a specific query, it is possible that only data from a subset of users are important; in this case, it would be beneficial to exclude other users from participation and allow them to save privacy budget. To this end, we assume there is a real-valued $1$-Lipschitz function $\phi(\cdot)$ which provides information about how useful a user's data is for computing a desired query, where $x_i$ is deemed more useful than $x_j$ if $\phi(x_i)<\phi(x_j)$. 
The Lipschitzness assumption allows us to privately estimate each $\phi(x_i)$ with a reasonable error bound. 
We develop two template algorithms depending on how $\phi$ is to be used. Specifically, we consider two settings: 1) $\phi$ provides an absolute measure, i.e., $\phi(x_i)$ alone is sufficient to determine whether user $i$ should be excluded; 2) $\phi$ provides a relative measure, and we need to consider $\phi(x_1),\dotsb,\phi(x_n)$ together to determine which users should be excluded. We refer to the two settings as non-interactive and interactive elimination, respectively. 

\subsubsection{Private Non-interactive Iterative Elimination}
Suppose we have $\nu_{\mathrm{low}} \le \nu_{\mathrm{high}}\in \mathbb{R}\cup \{\pm \infty\}$ such that $x_i$ is considered useful for computing the query if $\phi(x_i)< \nu_{\mathrm{low}}$, and not useful if $\phi(x_i)> \nu_{\mathrm{high}}$. Here, we develop an algorithm to find two sets of indices $S_0, S_1$ so that with high probability, $\phi(x_i) < \nu_{\mathrm{low}}$ if $i\in S_1$, and $\phi(x_i) > \nu_{\mathrm{high}}$ if $i\in S_0$. 
To this end, we further assume that there is a triple $(M, g, h)$ where:
\begin{enumerate}
    \item $M(\cdot)$ privatizes (some function of) $x_i$; 
    \item $g(\cdot)$ estimates $\phi(x_i)$ from $M(x_i)$ (possibly with post-processing);
    \item $h(\cdot)$ computes an error bound on the estimate of $\phi(x_i)$.
\end{enumerate}
\begin{definition}
    Let $\phi(\cdot)$ be a $1$-Lipschitz function. Fix any $c\in \mathbb{Z}_{>0}$. For $1\le j \le c$, let $r_j>0$ and $\tilde{v}(j):=M(u,r_{j})$ be the privatized output obtained by applying $M$ to $u$ with privacy parameter $r_{j}$. Let $\bar{\phi}(j):=g(\{\tilde{v}(s)\}_{s\le j})$ be an estimate of $\phi(u)$ obtained from applying $g(\cdot)$ to the privatized outputs $\{\tilde{v}(s)\}_{s\le j}$. If for each $j\in [c]$,  any $\beta\in(0,1)$, and arbitrary $u$ it holds with probability $1-\beta$ that
$
|\bar{\phi}(j)-\phi(u)|\le h(\{r_s\}_{s\le j},\beta),
$
 then we call $(M,g,h)$ a valid triple for $\phi(\cdot)$.
\end{definition}
\begin{algorithm}[htbp]
    \caption{Private Iterative Elimination (Non-interactive): $\mathrm{PIE}$-$\mathrm{NI}$}
    \label{algo:pie_non}
        \begin{flushleft} 
        \textbf{Input}: 
        $x_1,\dotsb,x_n$; $G_0\subseteq [n]$; number of rounds $c\in\mathbb{Z}_{> 0}$;  $\beta_0 > 0$; $r_{1,[c]},\dotsb,r_{n,[c]} > 0$; $1$-Lipschitz $\phi$; private algorithm $M$; estimation function $g$; width function $h$ computes bound for $g$; $\nu_{\mathrm{low}}\le\nu_{\mathrm{high}}\in \mathbb{R}\cup\{\pm \infty\}$; privacy budgets $B_{[n]}$\\
        \textbf{Output}: 
        subsets of indices $S_0, S_1, G$; collection of privatized outputs $\{\tilde{v}_i\}_{i\in G_0}$
        \end{flushleft}
        \begin{algorithmic}[1]
        \State $S_0 \gets \emptyset; \;\;  S_1\gets \emptyset; \;\; j \gets 1$
        \While{$j \le c$ and $|G_{j-1}|>0 $}
        \State $G_{j}\gets \emptyset$
        \For{$i\in G_{j-1}$}
            \State $\tilde{v}_i(j)\gets M(x_i,r_{i,j})$
            \State $B_i \gets B_i - r_{i,j}$
            \State $\bar{\phi}_i(j)\gets g(\{\tilde{v}_i(s)\}_{s\le j};\phi)$ 
            \Comment{update estimate of $\phi(x_i)$}
            \State $\bar{h}_{i,j}\gets h(r_{i,[j]},\frac{\beta_0}{c|G_{j-1}|})$ \Comment{bounds $|\bar{\phi}_i(j)-\phi(x_i)|$}
        \If{$\bar{\phi}_i(j) < -\bar{h}_{i,j}+\nu_{\mathrm{low}}$}
        \State $S_1 \gets S_1 \cup \{i\}$
        \ElsIf{$\bar{\phi}_i(j) > \bar{h}_{i,j}+\nu_{\mathrm{high}}$}
        \State $S_0 \gets S_0 \cup \{i\}$
        \Else
        \State $G_{j}\gets G_{j}\cup \{i\}$
        \EndIf
        \EndFor
        \State $j \gets j+1$
        \EndWhile
        \State output $S_0, S_1, G_{j-1}, \{\tilde{v}_i\}_{i\in G_0}$ \Comment{lengths of $\tilde{v}_i$'s vary} 
        \end{algorithmic}
    \end{algorithm}
Our algorithm proceeds in iterations as follows, after we split the assigned budget $r_i$ into $c$ parts: We start with the set $G_0\subseteq[n]$ to be checked, use one part of $r_i$ to obtain a noisy estimate of $\phi(x_i)$, for each $i$. At each iteration $j$, we check if the noisy estimate $\bar{\phi}_i(j)$ is sufficiently far from $[\nu_{\mathrm{low}},\nu_{\mathrm{high}}]$, and if so, $i$ is eliminated from the remaining iterations.
We check until at most $c$ iterations or if no elements remain in $G_j$. 
The details are given in Algorithm~\ref{algo:pie_non}, which produces the sets $S_0$ and $S_1$ with the following guarantee:
\begin{lemma}
\label{lm:pie_non_utility}
In Algorithm~\ref{algo:pie_non}, let $(M,g,h)$ be a valid triple for $\phi(\cdot)$. 
Then with probability $1-\beta_0$, we have simultaneously that 
\[\forall i\in S_1: \phi(x_{i})<\nu_{\mathrm{low}} \mathrm{\;and\;} \forall i\in S_0: \phi(x_{i})>\nu_{\mathrm{high}}.\]
\end{lemma}

After we obtain the sets $S_0, S_1$ and the collection of privatized outputs $\{\tilde{v}_i\}_{i\in G_0}$, what we do with them depends on the specific application, which will be discussed in Section~\ref{sec:adagp_app}.

\subsubsection{Private Interactive Iterative Elimination}
In the interactive case, to determine whether a user $i$ should be excluded, we need to determine the value of $\phi(x_i)$ \textit{relative} to that of $\phi(x_j)$ for all $j\neq i$. Specifically, we consider the setting where a user $i$ should be excluded if $\phi(x_i)$ is not among the $k$ smallest such values, for a fixed $k\ge 1$. Here, we develop an algorithm to filter out such users. 
\begin{algorithm}[htbp]
    \caption{Private Iterative Elimination (Interactive): $\mathrm{PIE}$-$k$}
    \label{algo:pie_topk}
    \begin{flushleft}
        \textbf{Input}: $x_1,\dotsb,x_n$; $G_0\subseteq [n]$ ; $k,c\in \mathbb{Z}_{> 0}$; $r_{1,[c]},\dotsb,r_{n,[c]} > 0$; $\beta_0 > 0$; $1$-Lipschitz $\phi$; private algorithm $M$; estimation function $g$; width function $h$ computes bound for $g$; privacy budgets $B_{[n]}$\\
        \textbf{Output}: subset of indices $G$; collection of privatized outputs $\{\tilde{v}_i\}_{i\in G_0}$
     \end{flushleft}
        \begin{algorithmic}[1]
        \State $j \gets 1$
        \While{$j \le c$ and $|G_{j-1}|>k$}
        \For{$i\in G_{j-1}$}
            \State $\tilde{v}_i(j) \gets M(x_i,r_{i,j})$ 
            \State $B_i \gets B_i - r_{i,j}$
            \State $\bar{\phi}_i(j) \gets g\left(\{\tilde{v}_i(s)\}_{s\le j};\phi \right)$ 
        \State $\bar{h}_{i,j}\gets h(r_{i,[j]},\frac{\beta_0}{c|G_{j-1}|})$
        \EndFor
        \State $(t_1,\dotsb,t_k) \gets \min_{S\subset G_{j-1}:|S|=k} \{\sum_{i\in S}\bar{\phi}_i(j)+\bar{h}_{i,j}\}$ 
        \State $I_{t_k}(j) \gets [\bar{\phi}_{t_k}(j) - \bar{h}_{t_k,j}, \bar{\phi}_{t_k}(j) + \bar{h}_{t_k,j}]$
        \State $G_{j}\gets \{t_1,\dotsb,t_k\}$
        \For{$i\in G_{j-1}$}
        \State $I_{i}(j) \gets [\bar{\phi}_i(j) - \bar{h}_{i,j}, \bar{\phi}_i(j) + \bar{h}_{i,j}]$
        \If{$I_{i}(j)\cap I_{{t_k}}(j) \neq \emptyset$}
        \State $G_{j}\gets G_{j}\cup \{i\}$
        \EndIf
        \EndFor
        \State $j \gets j+1$
        \EndWhile
        \State output $G_{j-1}, \{\tilde{v}_i\}_{i\in G_0}$ \Comment{lengths of $\tilde{v}_i$'s vary}  
        \end{algorithmic}
    \end{algorithm}
    
We again assume that there is a triple $(M,g,h)$ for estimating $\phi(\cdot)$ and computing an error bound for the estimate, as introduced in the non-interactive case.  
However, unlike the non-interactive case where the noisy estimate of $\phi(x_i)$ is compared against pre-specified values, here we need to compare noisy estimates of the $\phi(x_i)$'s against each other. When a small privacy parameter is used for privatizing $\phi(x_i)$, the resulting estimate might largely deviate from the true value. Thus, we will compare based on the confidence intervals of these noisy estimates.

Our algorithm is given in Algorithm~\ref{algo:pie_topk} and proceeds as follows: We start with the set $G_0\subseteq \{1,2,\dotsb,n\}$;  
for each $x_i$ we use a part of $r_i$ to get a privatized estimate of $\phi(x_i)$ (line 4-6). At each iteration $j$, we compute the width $\bar{h}_{i,j}$ of the confidence interval $I_i(j)$ for $\phi(x_i)$, for each user $i$ (line 7). Next, we compare the noisy values of the $\phi(x_i)$'s based on the right end points of their confidence intervals; specifically, we determine among the subset $G_{j-1}$ the $k$ points $\{t_1,\dotsb,t_k\}$ with the smallest right end points (line $8$).
Then, we determine the points with noisy values sufficiently close to that of user $t_k$, based on whether their confidence intervals overlap with $I_{t_k}(j)$ (line $9$-$14$); these points together with $\{t_1,\dotsb,t_k\}$ form the subset $G_j$ that will be further checked at the next iteration, while all other points are eliminated from further inspection. We continue the process for $c$ iterations or until only $k$ points remain. 
This elimination process has the following guarantee:
\begin{lemma}   
\label{lm:J_subset}
In Algorithm~\ref{algo:pie_topk}, let $(M,g,h)$ be a valid triple for $\phi(\cdot)$.
    Let $J^*:=\{i^*_1, \dotsb,i^*_k\}$ denote the set of indices corresponding to the points with the $k$ smallest values when evaluated by $\phi(\cdot)$. Let $\hat{j}$ be the last iteration before the while-loop exits. Then, with probability $1-\beta_0$, $J^*$ is a subset of every $G_j$ where $j\le \hat{j} \le c$.
\end{lemma}

\section{Applications}
\label{sec:adagp_app}
In this section, we consider three applications in the local model setting where the algorithms developed above can be applied. As stated in the previous section, the algorithms apply to a specified component of each $x_i$. To avoid clutter of notation, we just write $x_i$ to mean $x_i(l)$ in some component $U_l$ to be queried. We further assume $U_l\subseteq \mathbb{R}^d$ is equipped with the Euclidean metric $\|\cdot\|$ for the applications in this section. For the purpose of demonstrating our techniques, we consider privatization under CGP.

In Section~\ref{sec:rc}, we consider the range counting query where we estimate the number of points inside a range $\Gamma$; in Section~\ref{sec:gauss_kde}, we compute the value of a Gaussian kernel density estimator (KDE) function at a given point $p$;
in Section~\ref{sec:knn}, we find the $k$ nearest neighbors ($k$NN) among the $x_i$'s to a given point $p$. 
Recall that Lemma~\ref{lm:basic_mech} provides basic $\rho$-CGP mechanisms of the form $M(\cdot,\rho;\varphi): u\mapsto \varphi(u)+\frac{K}{\sqrt{2\rho}}\cdot Z$, where $Z\sim \mathcal{N}(0,I_{d\times d})$, for $K$-Lipschitz function $\varphi$ with range $V\subseteq\mathbb{R}^d$; by choosing the function $\varphi$, we get different baseline solutions for the queries mentioned above, where we apply $M(\cdot,\rho;\varphi)$ to each $x_i$, then aggregate the privatized outputs after applying appropriate post-processing. Given such a basic mechanism $M$,
Algorithms~\ref{algo:pie_non} and \ref{algo:pie_topk}
allow us to turn it into one that allows privacy savings, and we discuss how to instantiate the algorithms for each of the mentioned queries in the subsections below.

\subsection{Technical Lemmas}
In order to apply Algorithm~\ref{algo:pie_non} or \ref{algo:pie_topk}, we need a function $\phi(\cdot)$ for determining whether each $x_i$ is useful for computing the desired query, and a triple $(M,g,h)$ for privately estimating $\phi(\cdot)$. The function $\phi(\cdot)$ will be discussed in the subsections following this; here, we state some results on Gaussian noises which are needed to construct the functions $g$ and $h$.
For bounding Gaussian noises, we have:
\begin{lemma} [Gaussian Bounds \cite{laurent2000adaptive, liang2023concentrated}]
\label{lm:upper_bound_noise}
Let $Z\sim\mathcal{N}\left(0,{I}_{d\times d}\right)$ where $d>2$. Then with probability at least $1-\beta$, we have 
\[
\|Z\| \leq \sqrt{d+2\sqrt{d\log(1/\beta)}+2\log(1/\beta)}=:\lambda(d,\beta).
\]
For $d=2$ we have $\|Z\|\le \sqrt{2\log({1}/{\beta})}=:\lambda(2,\beta)$, and for $d=1$ we have $|Z|\le \sqrt{2\log(2/\beta)}=:\lambda(1,\beta)$.
\end{lemma}

Suppose we split $\rho$ into $c$ parts so that $\sum_{j\in[c]}\rho_j = \rho$, and let $v_j:=\frac{1}{\sqrt{2\rho_j}}Z_j$ where $Z_j\sim\mathcal{N}(0,I_{d\times d})$ for $j\in [c]$.
The variance-weighted average $\bar{v}:=\sum_{j\in[c]}({\rho_j}/{\rho}) v_j$ of the $v_j$'s is distributed as $\mathcal{N}(0,({1}/{2\rho})I_{d\times d})$.
In particular, the partial averages have the following distributions, where $\bar{\rho}_j:={\sum}_{s\le j}\rho_s$:

\begin{fact} [Linear combinations of  Gaussian distribution]
\label{ft:lincomb_gauss}
Fix $c\in \mathbb{Z}_{>0}$. For $1\le j\le c$, let $\rho_j>0$, $Z_j\in \mathcal{N}(0, I_{d\times d})$ and and assume the $Z_j$'s are independent.  Then the random variable $Y_j:=\frac{1}{\bar{\rho}_j}\sum_{s=1}^{j}\rho_s\frac{1}{\sqrt{2 \rho_s}} Z_s$ follows $Y_j\sim \mathcal{N}\left(0, ({1}/{2\bar{\rho}_j}) {I}_{d\times d}\right)$. 
\end{fact}

\begin{corollary}
\label{cor:g0h0}
    Fix any $1$-Lipschitz $\phi(\cdot)$. 
    For $u\in \mathbb{R}^d$, $j\in [c]$, let $v_j:=u+\frac{1}{\sqrt{2\rho_j}} Z_j$ where $Z_j\sim \mathcal{N}\left(0,I_{d\times d}\right)$. Let  
    $g_0:\left(\{v_s\}_{s\le j} \right)\mapsto \phi\left(\frac{1}{\bar{\rho}_j}\sum_{s\le j}{\rho_s}v_s\right)$, 
    and $h_0:\left(\{\rho_s\}_{s\le j},\beta\right)\mapsto \lambda(d,\beta)/{\sqrt{2\bar{\rho}_j}}$.
    Then with probability $1-\beta$,
    \begin{align*}
    \left|g_0(\{v_s\}_{s\le j})-\phi(u)\right|&\le \left\|\frac{1}{\bar{\rho}_j}\sum\nolimits_{s\le j}(\rho_s v_s)-u\right\|
    \le h_0\left(\{\rho_s\}_{s\le j},\beta\right).
    \end{align*}
\end{corollary}
Thus, $(M(\cdot,\cdot;f_{\mathrm{id}}),g_0,h_0)$ is a valid triple for $\phi(\cdot)$, where $f_{\mathrm{id}}:u\mapsto u$ is the identity function. 

\begin{corollary}
\label{cor:g1h1}
    Fix any $1$-Lipschitz $\phi(\cdot)$. 
    For $u\in \mathbb{R}^d$,
    let $v_j:=\phi(u)+\frac{1}{\sqrt{2\rho_j}}Z_j$ where $Z_j\sim \mathcal{N}\left(0,1\right)$ for $j\in [c]$. Let $g_1:\{v_s\}_{s\le j} \mapsto \left(\frac{1}{\bar{\rho}_j}\sum_{s\le j}\rho_s v_s\right)$. Let $h_1:\left(\{\rho_s\}_{s\le j},\beta\right)\mapsto \lambda(1,\beta)/{\sqrt{2\bar{\rho}_j}}$. 
    Then with probability $1-\beta$,
    \[
    \left|g_1\left(\{v_s\}_{s\le j}\right)-\phi(u)\right| \le h_1\left(\{\rho_s\}_{s\le j},\beta\right).
    \]
\end{corollary}
Thus, $(M(\cdot,\cdot;\phi),g_1,h_1)$ is a valid triple for $\phi(\cdot)$.
\begin{remark}
    In the above, the mechanism $M(\cdot,\cdot;f_{\mathrm{id}})$ for privatizing $f_{\mathrm{id}}:u\mapsto u$ requires the use of Euclidean metric, while $M(\cdot,\cdot;\phi)$ works for an arbitrary metric; in particular, the latter can be used with the Hamming metric, as demonstrated in the central model application in Appendix~\ref{app:cdpcount}.
\end{remark}

\subsection{Range Counting}
\label{sec:rc}
Suppose there is a range $\Gamma\subset U_l\subseteq\mathbb{R}^d$. For a collection of points $x_1,\dotsb,x_n$, we are interested in counting how many points fall within the range. 
The exact number can be computed as 
\[f(x_1,\dotsb,x_n; \Gamma)=\sum\nolimits_{i} \mathbb{1}_{\Gamma}(x_{i}),\] 
where $\mathbb{1}_{\Gamma}:u\rightarrow \{0,1\}$ and $\mathbb{1}_{\Gamma}(u)=1$ only if $u\in \Gamma$. Another way to compute $f(x_1,\dotsb,x_n; \Gamma)$ is by checking the distances of each point to the boundary $\partial \Gamma$.
Denote the boundary of $\Gamma$ by $\partial\Gamma$. Consider a function $\ProjG:U_l\rightarrow \mathbb{R}$ defined by
\begin{align}
\label{eqn:proj_def}
\ProjG(u)=\begin{cases}
\;\;\;\min_{z\in \partial\Gamma} \|u-z\|,  &\text{if } u\notin \Gamma\\
-\min_{z\in \partial\Gamma} \|u-z\|,  &\text{if } u\in \Gamma.
\end{cases}
\end{align}
Then we can count the number of points inside $\Gamma$ by computing $\sum_i \mathbb{1}\{\ProjG(x_i)<0\}$.
\begin{lemma} 
\label{lm:projgamma_lipschitz}
For a range $\Gamma$ enclosed by a closed curve $\partial\Gamma$, the function $\ProjG$ defined in equation~\eqref{eqn:proj_def} is $1$-Lipschitz.
\end{lemma}

\subsubsection{Private Range Counting: Baseline Methods}
\label{sec:rc_baseline}
\paragraph{Point Privatization.}
  To compute $f(x_1,\dotsb,x_n; \Gamma)$ privately, one immediate solution is to first privatize the points, then count the privatized points; i.e., 
we compute $\tilde{x}_i=M(x_i,\rho_i;f_{\mathrm{id}})=x_i+\frac{1}{\sqrt{2\rho_i}} Z_i$,  $Z_i\sim\mathcal{N}(0,I_{d\times d})$ for $i\in [n]$, then apply $\mathbb{1}_{\Gamma}(\cdot)$ to each $\tilde{x}_i$ to get
\begin{align*}
\mathrm{BM}_0(x_1,\dotsb,x_n)&:=f(\tilde{x}_1,\dotsb,\tilde{x}_n;\Gamma)\\
&=\sum\nolimits_{i\in[n]} \mathbb{1}_{\Gamma}(\tilde{x}_{i})=\sum\nolimits_{i\in[n]} \mathbb{1}\{\ProjG(\tilde{x}_i)<0\}.
\end{align*} 
By Lemma~\ref{lm:projgamma_lipschitz}, $|\ProjG(\tilde{x}_i)-\ProjG({x}_i)|\le \|\tilde{x}_i-x_i\|$ for each $i$. 
Thus, we have for all $i \in [n]$ simultaneously, with probability $1-\beta$
\[
|\ProjG(\tilde{x}_i)-\ProjG({x}_i)|\le \|\tilde{x}_i-x_i\|\le \lambda(d,\beta/n)/{\sqrt{2\rho_i}}.
\]
 Therefore, $\mathrm{BM}_0$ correctly counts all the points $x_i$ where $|\ProjG(x_i)|> {\lambda(d,\beta/n)}/{\sqrt{2\rho_i}} $ with probability $1-\beta$.

\paragraph{Distance Privatization.}
Another solution is to privatize $\ProjG(x_i)$ for each $x_i$, since $\ProjG$ is $1$-Lipschizt.
I.e., for $i\in [n]$, we compute $\tilde{y}_i=M(x_i,\rho_i;\ProjG)=\ProjG(x_i)+\frac{1}{\sqrt{2\rho_i}} Z_i$,  $Z_i\sim\mathcal{N}(0,1)$.
Then, we check if each noisy distance $\tilde{y}_i$ is near zero, specifically we compute 
\[
\mathrm{BM}_1(x_1,\dotsb,x_n):=\sum\nolimits_{i\in [n]} \mathbb{1}\{\tilde{y}_i<\eta_i\}
\]
for some suitable thresholds $\eta_i$'s. 
 We have for all $i\in [n]$ simultaneously, with probability $1-\beta$
\[
\left|\tilde{y}_i-\ProjG(x_i)\right| \le \lambda(1,\beta/n)/{\sqrt{2\rho_i}}.
\]
Thus, with probability $1-\beta$, $\mathrm{BM}_1$ correctly counts all the points $x_i$ where $|\ProjG(x_i)-\eta_i|>\frac{1}{\sqrt{2\rho_i}}\cdot \lambda(1,\beta/n)$. Note although we can simply set $\eta_i=0$ for all $i\in [n]$, we might end up with an over-estimate for $d>1$ if the privacy parameters $\rho_i$'s are small; this is because for a convex $\Gamma\subseteq \mathbb{R}^d$, the region near $\partial\Gamma$ has more volume on the outside than on the inside of $\Gamma$. We describe a method to mitigate this issue in Appendix~\ref{appendix:rc_thres_shift}, which sets $\eta_i=-{a_i}/{\sqrt{2\rho_i}}$, where $a_i$ is a small constant in $[0,1]$.
     Thus, $\mathrm{BM}_1$ is preferable to $\mathrm{BM}_0$ only when the suitable $\eta_i$'s could be easily calculated.

\subsubsection{Private Range Counting via Iterative Elimination}
Next, we discuss how to use iterative elimination to get privacy savings. For this query, we want to find $|I_\text{in}|$ where $I_\text{in}:=\{i\in[n]:\ProjG(x_i)<0\}$. Thus, a point $x_i$ with positive $\ProjG(x_i)$ is not useful for estimating $|I_\text{in}|$, and an obvious choice of $\phi(\cdot)$ is $\ProjG(\cdot)$, which is $1$-Lipschitz. Since we only need to check whether $\ProjG(\cdot)<0$, we can use non-interactive elimination (Algorithm~\ref{algo:pie_non}) with $\nu_{\mathrm{low}}=0=\nu_{\mathrm{high}}$.
By Corollaries~\ref{cor:g0h0} and \ref{cor:g1h1}, we have that both $(M(\cdot,\cdot;f_{\text{id}}),g_0,h_0)$ and $(M(\cdot,\cdot;\phi),g_1,h_1)$ are valid triples for $\ProjG(\cdot)$, corresponding respectively to the privatization mechanisms for $\mathrm{BM}_0$ and $\mathrm{BM}_1$.
\begin{algorithm}[h]
    \caption{Private Range Counting via Iterative Elimination}
    \label{algo:rc_pie_pm}
        \begin{flushleft} 
        \textbf{Input}: 
        $x=(x_1,\dotsb,x_n)\subset U_l$; $\Gamma\subset U_l$; $c\in \mathbb{Z}_{> 0}$;  $\beta_0 > 0$; $r=r_{1,[c]},\dotsb,r_{n,[c]} > 0$; valid triple $(M,g,h)$; privacy budget $B_{[n]}$\\
        \textbf{Output}: privatized count $\tilde{C}$
        \end{flushleft}
        \begin{algorithmic}[1]
        \State $S_0, S_1, G_{\hat{j}}, \{\tilde{v}_i\}_{i} \gets \mathrm{PIE}\text{-}\mathrm{NI}(x,[n],c,\beta_0,r,\ProjG, M,g,h,0,0, B_{[n]})$

        \If{$G_{\hat{j}} = \emptyset$}
            \State $\tilde{C} \gets |S_1|$ 
            \Comment{occurs if $\mathrm{PIE}\text{-}\mathrm{NI}$ completes $\hat{j}<c$ iterations}
        \Else
            \State $\tilde{C} \gets \mathrm{PM}(S_1,\{\tilde{v}_i\}_{i\in G_{c}})$ \Comment{postprocessing depends on $M$}
        \EndIf
        \State output $\tilde{C}$
        \end{algorithmic}
    \end{algorithm}
    
Suppose we run Algorithm~\ref{algo:rc_pie_pm} with the parameters specified above, where $\sum_{j\in[c]} r_{i,j} = \rho_i$ for each $i\in [n]$, and $S_0, S_1, G_{\hat{j}}, \{\tilde{v}_i\}_{i\in [n]}$ are the outputs of $\mathrm{PIE}\text{-}\mathrm{NI}$ (Algorithm~\ref{algo:pie_non}). Note $S_1$ and $S_0$ correspond to points where $\ProjG(\cdot)$ is less than or greater than zero, respectively. If $G_{\hat{j}} = \emptyset$, then $S_0\cup S_1=[n]$ and by Lemma~\ref{lm:pie_non_utility}:
 \begin{corollary}
\label{cor:prcie_ET_utility}
    Suppose $G_{\hat{j}} = \emptyset$ in Algorithm~\ref{algo:rc_pie_pm}. 
  Then, with probability $1-\beta_0$, all points are correctly categorized.
\end{corollary}
 That is, with probability $1-\beta_0$, $|S_1|=|I_\text{in}|$. Moreover, when $\hat{j}<c$, every $i\in [n]$ saves the remaining privacy budgets $r_{i,\hat{j}+1},\dotsb,r_{i,c}$, which can be used for later queries. 
 Next, consider the (bad) case where $\mathrm{PIE}\text{-}\mathrm{NI}$ completes $\hat{j}=c$ iterations.
 We need to apply a postprocessing step (line $5$ of Algorithm~\ref{algo:rc_pie_pm}) to compute the privatized count $\tilde{C}$, 
 which depends on the privatization function $M$: 

\paragraph{$\mathrm{PIE}$-$\mathrm{NI}$ with Point Privatization.} For $M=M(\cdot,\cdot;f_{\text{id}})$, 
 let 
 \[
\mathrm{PM}_0(S_1,\{\tilde{v}_i\}_{i\in G_{c}}):=|S_1|+{\sum}_{i\in G_c}\mathbb{1}\{\ProjG(\bar{x}_i)\},\]
where $\bar{x}_i:= \frac{1}{\rho_i}\sum_{j\in[c]}r_{i,j}\tilde{v}_i(j)$. By Fact~\ref{ft:lincomb_gauss}, $\bar{x}_i\sim \mathcal{N}(x_i,\frac{1}{2\rho_i}I_{d\times d})$, so with probability $1-\beta_1$, $\mathrm{PM}_0$ correctly counts all the points $x_i$ in $G_c$ where $|\ProjG(x_i)|>{\lambda(d,\beta_1/|G_c|)}/{\sqrt{2\rho_i}}$. 
 \begin{figure}[h]
     \centering
\includegraphics[width=0.32\textwidth]{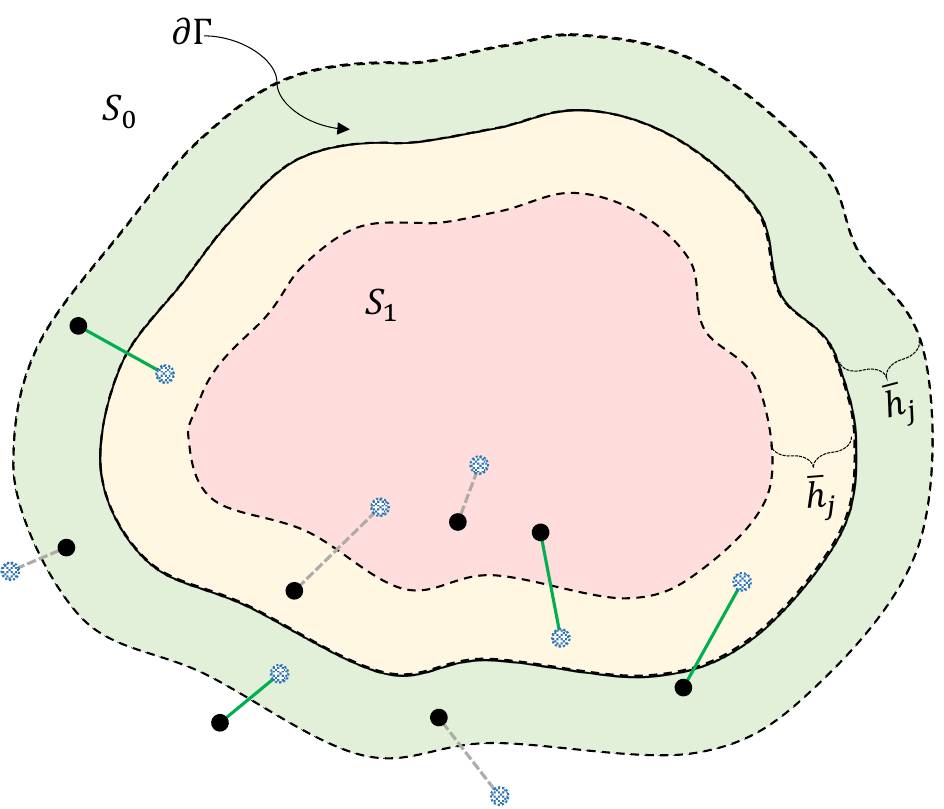}
            \label{fig:range_adv}
         \caption{Selection of points to $G_{j}$ in the call $\mathrm{PIE}$-$\mathrm{NI}$ of Algorithm~\ref{algo:rc_pie_pm} for range counting via point privatization. True locations: black solid dots; privatized locations: blue patterned dots. Solid green lines: points that advance to next round; gray dashed lines: points eliminated at this round. 
         % For the purpose of illustration here we assume all points have the same privacy budget so 
         Assume $\bar{h}_{i,j}=\bar{h}_j$. 
          % for all $i\in G_{j-1}$.
         }
\end{figure}
\paragraph{$\mathrm{PIE}$-$\mathrm{NI}$ with Distance Privatization.} For $M=M(\cdot,\cdot;\ProjG)$, let $\bar{y}_i:=\frac{1}{\rho_i}\sum_{j\in[c]}r_{i,j}\tilde{v}_i(j)$ be the weighted average of all $c$ estimates of $\ProjG(x_i)$. We estimate $|I_\text{in}|$ as 
\[
\mathrm{PM}_1(S_1,\{\tilde{v}_i\}_{i\in G_{c}}):=|S_1|+\sum\nolimits_{i\in G_c}\mathbb{1}\{\bar{y}_i<\eta_i\}\]
for the same $\eta_i$'s as in $\mathrm{BM}_1$ in Section~\ref{sec:rc_baseline}. By Fact~\ref{ft:lincomb_gauss}, $\bar{y}_i\sim\mathcal{N}(\ProjG(x_i),\frac{1}{2\rho_i})$, so $\mathrm{PM}_1$ correctly counts all $x_i$ in $G_c$ where with probability $1-\beta_1$, $|\ProjG(x_i)-\eta_i|> {\lambda(1,\beta_1/|G_c|)}/{\sqrt{2\rho_i}}$.

\begin{remark}
    In the methods $\mathrm{PM}_0$ and $ \mathrm{PM}_1$ above, set $\beta_0=0.25\beta$, $\beta_1=0.75\beta$, then $\mathrm{PM}_0, \mathrm{PM}_1$ are no worse than their respective baseline methods $\mathrm{BM}_0$, $\mathrm{BM}_1$ when at least $1/4$ of the points are eliminated in the iterative elimination process.
\end{remark}
 
\subsection{Gaussian KDE}
\label{sec:gauss_kde}
Fix $p\in U_l$. Given bandwidth $b>0$, we want to compute the Gaussian KDE estimate at $p$: $f(x_1,\dotsb,x_n;p,b):=\frac{1}{n}\sum_{i\in [n]} e^{-\frac{1}{2b^2}\|x_i-p\|^2}$. 
We have two baseline methods for privatizing the function value $f(x_1,\dotsb,x_n;p,b)$; one is from privatizing the $x_i$'s, and the other is from privatizing $\mathrm{len}_p(x_i)$, where $\mathrm{len}_p:u\mapsto \|u-p\|$ is $1$-Lipschitz. 

\subsubsection{Private KDE: Baseline Methods}
\label{sec:kde_baseline}
\paragraph{Point Privatization.} Let $\tilde{x}_i=M(x_i,\rho_i;f_\mathrm{id})$ 
 for $i\in [n]$. 
 Then $\tilde{x}_i\sim\mathcal{N}(x_i,\frac{1}{2\rho_i}I_{d\times d})$, 
 and we estimate $f(x_1,\dotsb,x_n;p,b)$ as
\[
\mathrm{BM}_0(x_1,\dotsb,x_n):=\frac{1}{n}\sum\nolimits_{i\in [n]}e^{-\frac{1}{2b^2}\mathrm{len}_p(\tilde{x}_i)^2}.
\]

\paragraph{Distance Privatization.}
Let $\tilde{y}_i=M(x_i,\rho_i;\mathrm{len}_p)$
for $i\in [n]$. 
Then $\tilde{y}_i\sim\mathcal{N}(\mathrm{len}_p(x_i),\frac{1}{2\rho_i})$, and we estimate the KDE value as
\[
\mathrm{BM}_1(x_,\dotsb,x_n):=\frac{1}{n}\sum\nolimits_{i\in [n]} e^{-\frac{1}{2b^2}\tilde{y}_i^2}.
\]

\subsubsection{Private KDE via Iterative Elimination}
For a point $x_i$ with large $\mathrm{len}_p(x_i)$, $e^{-\mathrm{len}_p(x_i)^2/(2b^2)}$ is small. E.g., if $\mathrm{len}_p(x_i)\ge 6 b$, then the point $x_i$ contributes at most ${e^{-18}}/{n}$ to $f(x_1,\dotsb,x_n;p,b)$. Thus, a natural choice for $\phi(\cdot)$ is $\mathrm{len}_p(\cdot)$, which means we eliminate the points that are far from $p$, and keep the near points. In this case, we are interested in a one-sided non-interactive elimination (Algorithm~\ref{algo:pie_non}); in particular, we set 
 $\nu_{\mathrm{low}}=-\infty$ in $\mathrm{PIE}\text{-}\mathrm{NI}$ which gives $S_1=\emptyset$. We have two valid triples $(M(\cdot,\cdot;f_{\mathrm{id}}),g_0,h_0)$ and $(M(\cdot,\cdot;\mathrm{len}_p),g_1,h_1)$ from Corollaries~\ref{cor:g0h0}, and \ref{cor:g1h1}, corresponding to the baseline methods above.
 \begin{algorithm}[h]
    \caption{Private KDE via Iterative Elimination}
    \label{algo:kde_pie_pm}
        \begin{flushleft} 
        \textbf{Input}: 
        $x=(x_1,\dotsb,x_n)\subset U_l$; $b>0$; $c\in \mathbb{Z}_{> 0}$;  $\beta_0 > 0$; $r=r_{1,[c]},\dotsb,r_{n,[c]} > 0$; valid triple $(M,g,h)$; privacy budget $B_{[n]}$\\
        \textbf{Output}: privatized KDE estimate $\tilde{A}$
        \end{flushleft}
        \begin{algorithmic}[1]
        \State $S_0, S_1, G_{\hat{j}}, \{\tilde{v}_i\}_{i} \gets \mathrm{PIE}\text{-}\mathrm{NI}(x,[n],c,\beta_0,r,\ProjG, M,g,h,-\infty,6b, B_{[n]})$
        \If{$G_{\hat{j}} = \emptyset$}
            \State $\tilde{A} \gets 0$ \Comment{occurs if $\mathrm{PIE}\text{-}\mathrm{NI}$ completes $\hat{j}<c$ iterations}
        \Else
            \State $\tilde{A} \gets \mathrm{PM}(S_0,\{\tilde{v}_i\}_{i\in G_{c}})$ \Comment{postprocessing depends on $M$}
        \EndIf
        \State output $\tilde{A}$
        \end{algorithmic}
    \end{algorithm}

Suppose we run Algorithm~\ref{algo:kde_pie_pm}, where $\sum_{j\in[c]} r_{i,j} = \rho_i$ for each $i\in [n]$. Let $S_0,S_1,G_{\hat{j}},\{\tilde{v}_i\}_{i\in [n]}$ be the outputs of {$\mathrm{PIE}\text{-}\mathrm{NI}$. 
By Lemma~\ref{lm:pie_non_utility}, every point $x_i$ in $S_0$ has $\mathrm{len}_p(x_i)>6b$, so every point in $S_0$ contributes at most $\frac{e^{-18}}{n}$ to the KDE estimate. If $G_{\hat{j}}=\emptyset$, then $S_0=[n]$ since $S_1=\emptyset$. Thus, we have:
\begin{corollary}
\label{cor:kde_ET_util}
    Suppose $G_{\hat{j}}=\emptyset$ in Algorithm~\ref{algo:kde_pie_pm}. Then with probability at least $1-\beta_0$, 
    \[\frac{1}{n}\sum_{i\in [n]} e^{-\frac{1}{2b^2}\|x_i-p\|^2} \le e^{-18}.\]
\end{corollary}

In this case, we simply estimate $f(x_1,\dotsb,x_n;p,b)$ as $\tilde{A}=0$ (line $3$ in Algorithm~\ref{algo:kde_pie_pm}), so the total error incurred is at most $e^{-18}$. If $G_{c} \neq \emptyset$, we apply the following postprocessing step (line $5$ of Algorithm~\ref{algo:kde_pie_pm}) depending on the privatization function $M$:

 \paragraph{$\mathrm{PIE}$-$\mathrm{NI}$ with Point Privatization.} 
For $M=M(\cdot,\cdot;f_{\mathrm{id}})$, Let
 \[
 \mathrm{PM}_0(S_0,\{\tilde{v}_i\}_{i\in G_{c}}):=\frac{1}{n}\sum\nolimits_{i\in G_c}e^{-\frac{1}{2b^2}\mathrm{len}_p(\bar{x}_i)^2}+\frac{|S_0|e^{-18}}{n},
 \]
where $\bar{x}_i:= \frac{1}{\rho_i}\sum_{j\in[c]}r_{i,j}\tilde{v}_i(j)\sim \mathcal{N}(x_i,\frac{1}{2\rho_i}I_{d\times d})$ for $i\in G_c$.

 \paragraph{$\mathrm{PIE}$-$\mathrm{NI}$ with Distance Privatization.} For $M=M(\cdot,\cdot;\mathrm{len}_p)$, let
 \[
\mathrm{PM}_1(S_0,\{\tilde{v}_i\}_{i\in G_{c}}):=\frac{1}{n}\sum\nolimits_{i\in G_c}e^{-\frac{1}{2b^2}\bar{y}_i^2} +\frac{|S_0|e^{-18}}{n},
 \]
where $\bar{y}_i:=\frac{1}{\rho_i}\sum_{j\in[c]}r_{i,j}\tilde{v}_i(j)
 \sim \mathcal{N}(\mathrm{len}_p(x_i),\frac{1}{2\rho_i})$ for $i\in G_c$.

  In the methods using iterative elimination above, the additional bias term (second term on RHS) may be omitted if it's larger than the first term; in any case, the additional error due to the elimination process is at most $e^{-18}$. 
  
\subsection{$k$ Nearest Neighbors Query}
\label{sec:knn}
Fix $k\in\mathbb{Z}_{>0}$ where $k<n$ and $p\in U_l$. For a collection of points $\{x_1,\dotsb,x_n\}$, the $k$ nearest neighbors query ($k$NN) computes the set of indices $\{i^*_1,\dotsb,i^*_k\}$ corresponding to the $k$ nearest points to $p$. I.e., 
$f(x_1,\dotsb,x_n;p):=\argmin_{S\subset [n]:|S|=k}\sum_{i\in S}\|x_i-p\|$. 

Similar to the case of range counting, there are two simple baseline methods for this query, which can be converted to privacy saving ones without compromising the utility. 

\subsubsection{Private $k$NN: Baseline Methods}
\paragraph{Point Privatization.}
Let $\tilde{x}_i=M(x_i,\rho_i;f_\mathrm{id})$ 
 for $i\in [n]$. Let
\begin{align*}
\label{eqn:knn_privloc}
\mathrm{BM}_0(x_1,\dotsb,x_n)
&:=\argmin_{S\subset [n]:|S|=k} \sum\nolimits_{i\in S}\|\tilde{x}_i-p\|.
\end{align*}
\begin{lemma}
\label{lm:knn_BM0}
Let $\{t_1,\dotsb,t_k\}$ denote the output computed from $\mathrm{BM}_0(x_1,\dotsb,x_n)$ defined above, where $\|\tilde{x}_{t_s}-p\|\le \|\tilde{x}_{t_r}-p\|$ for all $s\le r\in[k]$. 
For any $\beta > 0$, with probability at least $1-\beta$, we have simultaneously for all $s\in [k]$, where $\rho_0:=\min_{i\in [n]}\rho_i$,
\[
\|x_{t_s}-p\|-\|x_{i^*_s}-p\| \le \frac{\sqrt{2}}{\sqrt{\rho_0}}\cdot\lambda(d,\beta/n),
\]
where $\lambda(d,\beta/n)$ is given in Lemma~\ref{lm:upper_bound_noise}.
\end{lemma}
{\paragraph{Distance Privatization.}
Let $\tilde{y}_i=M(x_i,\rho_i;\mathrm{len}_p)$
for $i\in [n]$, where $\mathrm{len}_p:u\mapsto \|u-p\|$. Let
\begin{equation*}
\label{eqn:knn_privdist}
\mathrm{BM}_1(x_1,\dotsb,x_n)=\argmin_{S\subset [n]:|S|=k} \sum\nolimits_{i\in S} \tilde{y}_i.
\end{equation*}
\begin{lemma}
\label{lm:knn_BM1}
Let $\{t_1,\dotsb,t_k\}$ denote the output computed from $\mathrm{BM}_1(x_1,\dotsb,x_n)$ as defined above, where $\tilde{y}_{t_s}\le \tilde{y}_{t_r}$ for all $s\le r\in[k]$. 
For any $\beta > 0$, with probability $1-\beta$, we have simultaneously for all $s\in [k]$, where $\rho_0:=\min_{i\in [n]}\rho_i$,
\[
\|x_{t_s}-p\|-\|x_{i^*_s}-p\| \le \frac{\sqrt{2}}{\sqrt{\rho_0}}\cdot {\lambda(1,\beta/n)}.
\]
\end{lemma}
\subsubsection{Private $k$NN via Iterative Elimination} For this query, since we need to compare the $x_i$'s based on their distances to $p$, we will use interactive elimination with $\mathrm{len}_p(\cdot)$ as $\phi(\cdot)$. 
We again have two valid triples $(M(\cdot,\cdot;f_{\text{id}}),g_0,h_0)$ and $(M(\cdot,\cdot;\mathrm{len}_p),g_1,h_1)$ corresponding respectively to the baselines $\mathrm{BM}_0$ and $\mathrm{BM}_1$.
\begin{algorithm}[htbp]
    \caption{Private $k$NN via Iterative Elimination}
    \label{algo:knn_pie_pm}
        \begin{flushleft} 
        \textbf{Input}: 
        $x=(x_1,\dotsb,x_n)\subset U_l$; $p\in U_l$; $k,c\in \mathbb{Z}_{> 0}$; $\beta_0 > 0$; $r=r_{1,[c]},\dotsb,r_{n,[c]} > 0$; valid triple $(M,g,h)$; privacy buget $B_{[n]}$\\
        \textbf{Output}: set $\tilde{I}$ corresponding to $k$NN
        \end{flushleft}
        \begin{algorithmic}[1]
        \State $G_{\hat{j}}, \{\tilde{v}_i\}_{i\in [n]} \gets \mathrm{PIE}\text{-}k(x,[n],k,c,\beta_0,r,\mathrm{len}_p, M,g,h, B_{[n]})$
        \If{$|G_{\hat{j}}|=k$}
            \State $\tilde{I} \gets G_{\hat{j}}$ 
            \Comment{occurs if $\mathrm{PIE}\text{-}k$ completes $\hat{j}<c$ iterations}
        \Else
            \State $\tilde{I} \gets \mathrm{PM}(G_{c},\{\tilde{v}_i\}_{i\in G_{c}})$ \Comment{postprocessing depends on $M$}
        \EndIf
        \State output $\tilde{I}$
        \end{algorithmic}
    \end{algorithm}
    
Suppose we run Algorithm~\ref{algo:knn_pie_pm}, where $\sum_{j\in[c]} r_{i,j} = \rho_i$ for each $i\in [n]$,
and $G_{\hat{j}},\{\tilde{v}_i\}_{i\in[n]}$ are the outputs of $\mathrm{PIE}$-$k$. By Lemma~\ref{lm:J_subset}, $G_{\hat{j}}$ contains the $k$ points with the smallest $\mathrm{len}_p(\cdot)$ values, so if $|G_{\hat{j}}|=k$, then: 
 \begin{corollary}
 \label{cor:knn_ET_util}
    Suppose $|G_{\hat{j}}|=k$ in Algorithm~\ref{algo:knn_pie_pm}. 
    Then with probability $1-\beta_0$, $G_{\hat{j}}=\{i^*_1,\dotsb,i^*_k\}$.
\end{corollary}

If $|G_c|>k$, then we apply the following postprocessing (line $5$ of Algorithm~\ref{algo:knn_pie_pm}) to obtain $\tilde{I}$.

\paragraph{$\mathrm{PIE}$-$k$ with Point Privatization.} For $M=M(\cdot,\cdot;f_{\mathrm{id}})$, let
 \[
\mathrm{PM}_0(G_{c},\{\tilde{v}_i\}_{i\in G_{c}}):=\argmin_{S\subset G_c:|S|=k} \sum\nolimits_{i\in S}\|\bar{x}_i-p\|,
 \]
 where  $\bar{x}_i:= \frac{1}{\rho_i}\sum_{j\in[c]}r_{i,j}\tilde{v}_i(j)\sim \mathcal{N}(x_i,\frac{1}{2\rho_i}I_{d\times d})$ for $i\in G_c$.
\begin{lemma}
Let $(t_1,\dotsb,t_k)$ denote the output computed from $\mathrm{PM}_0$ above. Then with probability $1-(\beta_0+\beta_1)$,
\[
\|x_{t_s}-p\|-\|x_{i^*_s}-p\| \le \frac{\sqrt{2}}{\sqrt{\rho_0}}\cdot\lambda(d,\beta_1/|G_c|)
\]
simultaneously for all $s\in[k]$.
\end{lemma}
\begin{proof}
    By Lemma~\ref{lm:J_subset}, the set $\{i^*_1,\dotsb,i^*_k\}\subseteq G_c$ with probability $1-\beta_0$. Assume this holds, then Apply Lemma~\ref{lm:knn_BM0} with $G_c$ in place of $[n]$. The stated inequality holds with probability $1-(\beta_0+\beta_1)$ by a union bound.
\end{proof}
\paragraph{$\mathrm{PIE}$-$k$ with Distance Privatization.} For $M=M(\cdot,\cdot;\mathrm{len}_p)$, let
 \[
\mathrm{PM}_1(G_{c},\{\tilde{v}_i\}_{i\in G_{c}}):=\argmin_{S\subset G_c:|S|=k} \sum\nolimits_{i\in S}\bar{y}_i,
 \]
 where $\bar{y}_i:= \frac{1}{\rho_i}\sum_{j\in[c]}r_{i,j}\tilde{v}_i(j) \sim \mathcal{N}(\mathrm{len}_p(x_i),\frac{1}{2\rho_i})$ for $i\in G_c$.
 Then applying Lemmas ~\ref{lm:J_subset} and ~\ref{lm:knn_BM1}, and a union bound, we have
 \begin{lemma}
Let $(t_1,\dotsb,t_k)$ denote the output computed from $\mathrm{PM}_1$ above. Then with probability $1-(\beta_0+\beta_1)$,
\[
\|x_{t_s}-p\|-\|x_{i^*_s}-p\| \le ({\sqrt{2}}/{\sqrt{\rho_0}})\cdot\lambda(1,\beta_1/|G_c|)
\]
simultaneously for all $s\in[k]$.
\end{lemma}
In the methods $\mathrm{PM}_0$ and $\mathrm{PM}_1$ above, set $\beta_0=0.25\beta, \beta_1=0.75\beta$. Then $\mathrm{PM}_0, \mathrm{PM}_1$ are at least as good as their respective baseline methods $\mathrm{BM}_0, \mathrm{BM}_1$ when at least $1/4$ of the points are eliminated.

\section{Experiments}
\label{sec:experiments}
We conduct a set of experiments to evaluate the effectiveness of adaptive privacy budgeting on three real-world datasets: the MNIST dataset \cite{lecun1998}, the New York Motor Vehicle Collisions dataset (NYMVC) \cite{nyc2025mvc}, and the T-Drive trajectory dataset \cite{zheng2011t-drive}. For the location datasets, we convert the GPS coordinates to tuples in $\mathbb{R}^2$ using the Mercator projection,with meters ($\mathrm{m}$) as the unit of distance. 
We draw points $p_j$'s from the center area of approximately $100$km $\times$ $100$km for location datasets, and from the test dataset for MNIST;
 the $p_j$'s are used as query points, or as the centers for constructing the query ranges $\Gamma$. 

In each experiment, we evaluate the performance of the algorithms that use iterative elimination ($\mathrm{PM}_0, \mathrm{PM}_1$) against their respective baseline methods ($\mathrm{BM}_0$, $\mathrm{BM}_1$). The iterative algorithms involve splitting an allocated budget $\rho_i$ for user $i$ into $c$ parts, in the experiments we consider two ways of splitting $\rho_i$: 1) evenly splitting into $c$ parts so $\rho_i= \sum_{j\in[c]} \frac{\rho_i}{c}$; 2) doubling the proportion at each iteration, so $\rho_i = \sum_{j\in [c]} \frac{2^{j-1}}{2^c-1}\rho_i$. In the plots, we use the suffix \lq c4\rq, \lq c64\rq to denote the even splitting method with $c=4$ and $c=64$, respectively. The suffix \lq log-c10\rq corresponds to using doubling proportions with $c=10$.

\subsection{Single Query Privacy Savings}
We present experimental results to examine the potential privacy savings provided by the iterative algorithms in a single query, and to assess whether the elimination process in the iterative algorithms has an impact on the utility. 
Each experiment is repeated multiple ($\ge 300$) times, and we report the mean, the $25$-th and $75$-th percentiles of each error metric.  
In each experiment, the privacy saving of an algorithm $M$ is computed as:
\[
\mathrm{PrivSav}(M) = \frac{1}{n}\sum_{i\in[n]} {\hat{B}_i}/{\rho},
\]
where $\hat{B}_i$ represents remaining budget of user $i$ after $M$ is run, each user $i$ has an initial budget $\rho$.
\begin{figure}[h]
     \centering
\includegraphics[width=0.98\textwidth]{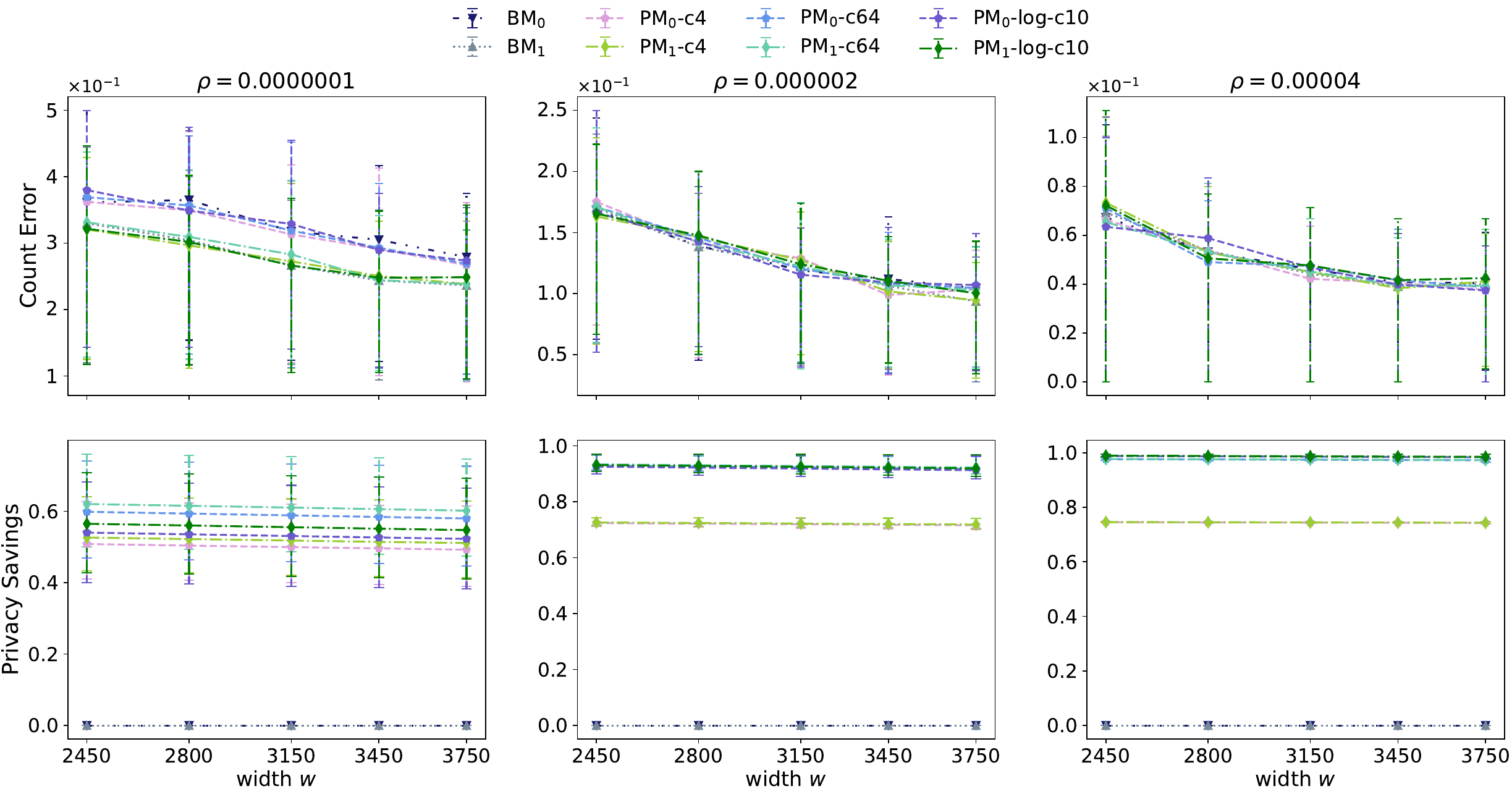}
         \caption{Single query privacy savings: range counting for various $w$ and privacy parameter $\rho$.}
        \label{fig:rc_1} 
\end{figure}

\paragraph{Single query: range counting} (Fig.~\ref{fig:rc_1})
We conduct the range counting experiment on the T-Drive dataset,  consisting of trajectories collected from approximately 10000 ($n=9740$) taxis in Beijing over a period of $7$ days. In each experiment, we sample an hour on a day, and square ranges $\Gamma_j$'s (see Appendix~\ref{appendix:projgamma} on how to compute $\ProjG$).
We sample three ranges per hour, for a total of $504=3\times 24\times 7$ repetitions. 
Let $C_j$ represent the true count of range $\Gamma_j$, and $\tilde{C}_j$ represent a privately computed count, for $j\in [504]$.
Then, we measure the relative count error as:
\[
\mathrm{CountErr}(\tilde{C}_j)={|\tilde{C}_j-C_j|}/{C_j}.\]

In Fig.~\ref{fig:rc_1} we observe similar performances in the $\mathrm{PM}$ algorithms compared to their $\mathrm{BM}$ counterparts. There is a slight advantage in the methods that use distance privatization over those that use point privatization when the privacy parameter is small. We see that privacy savings increase in the $\mathrm{PM}$ methods for larger privacy parameters, and large privacy savings occur even when $\rho$ is very small. The privacy savings are similar across the range widths $w$, since the same $\nu_{\mathrm{low}}=0=\nu_{\mathrm{high}}$ is used.

\paragraph{Single query: Gaussian KDE} (Fig.~\ref{fig:kde_1})
We perform the KDE experiment on the NYMVC dataset, where $n=200,000$ randomly drawn points are used in the experiments. The $p_j$'s are drawn from the vertices of a $200\times 200$ grid, for $j\in [300]$. 
Let $A_j$ represent represent the true KDE value at $p_j$, and $\tilde{A}_j$ represent a private estimate. Then we compute the $\ell_1$ error as: 
\[\ell_1\text{-}\mathrm{Err}(\tilde{A}_j)=|\tilde{A}_j-A_j|.\]
\begin{figure}[H]
     \centering
\includegraphics[width=0.98\textwidth]{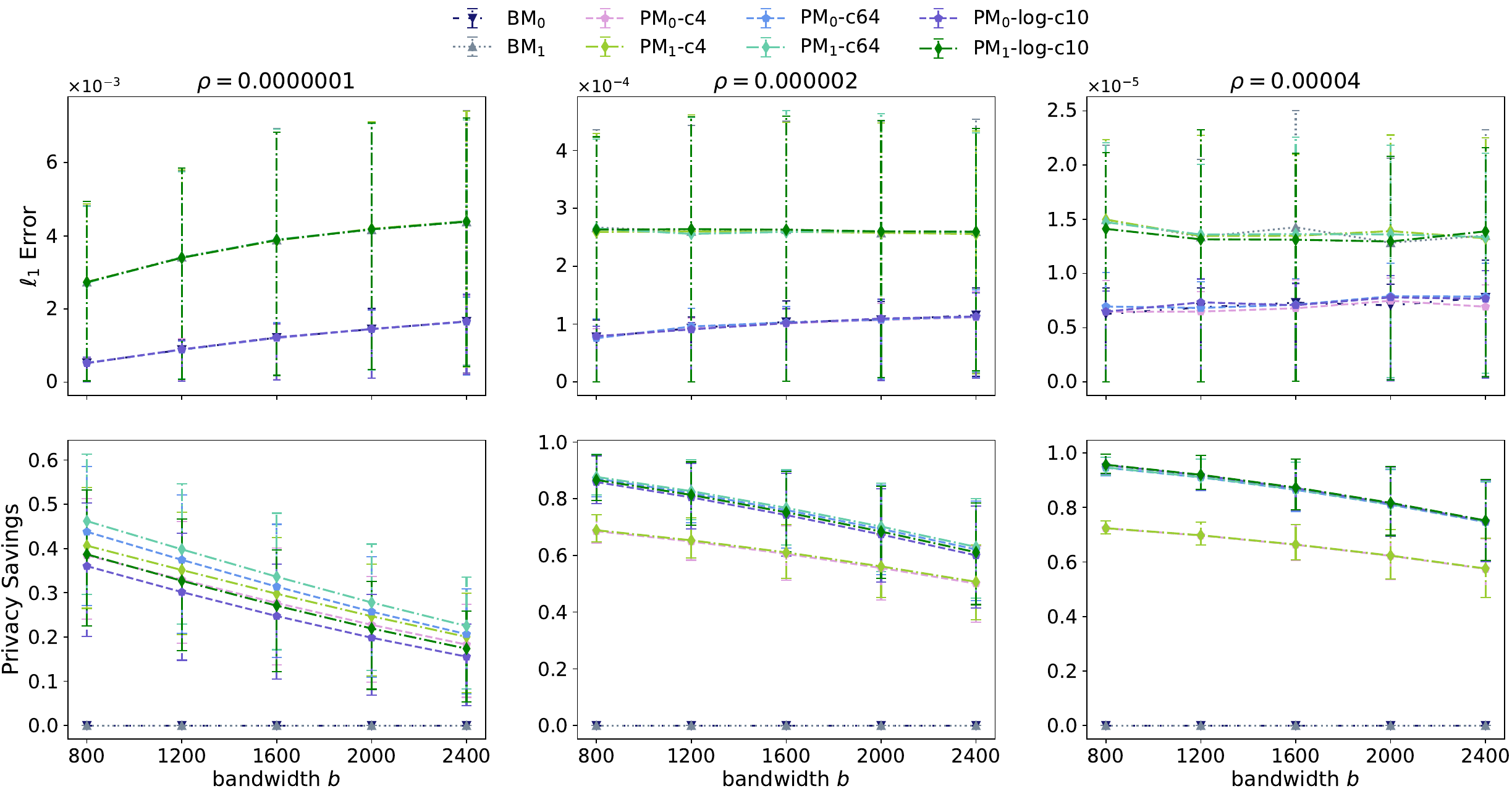}
         \caption{Single query privacy savings: KDE for various $b$ and privacy parameter $\rho$.}
        \label{fig:kde_1} 
\end{figure}

In Fig.~\ref{fig:kde_1}, we again observe similar performances in the $\mathrm{PM}$ vs. $\mathrm{BM}$ algorithms. Here, the methods using point privatization have an advantage over the methods that use distance privatization; this is due to the former incurring a smaller bias (negatively biased by as a scale of approximately $e^{-\|Z_2\|}$ for $Z_2\sim \mathcal{N}(0,I_{2\times 2})$) than the latter (positively biased by a scale of approximately $\mathbb{E}[e^{2Z_1\|x_i-p\|}]=O(e^{2\|x_i-p\|^2})$ where $Z_1\sim \mathcal{N}(0,1)$). We also see large privacy savings in the $\mathrm{PM}$ methods, which depend on the bandwidth $b$ in this case, since we set $\nu_{\mathrm{high}}=6b$. On the other hand, if there are many points farther than a distance of $6b$ from $p_j$, then the corresponding bandwidth $b$ might not be a good choice for KDE estimation.
\begin{figure}[htbp]
     \centering  
\includegraphics[width=0.98\textwidth]{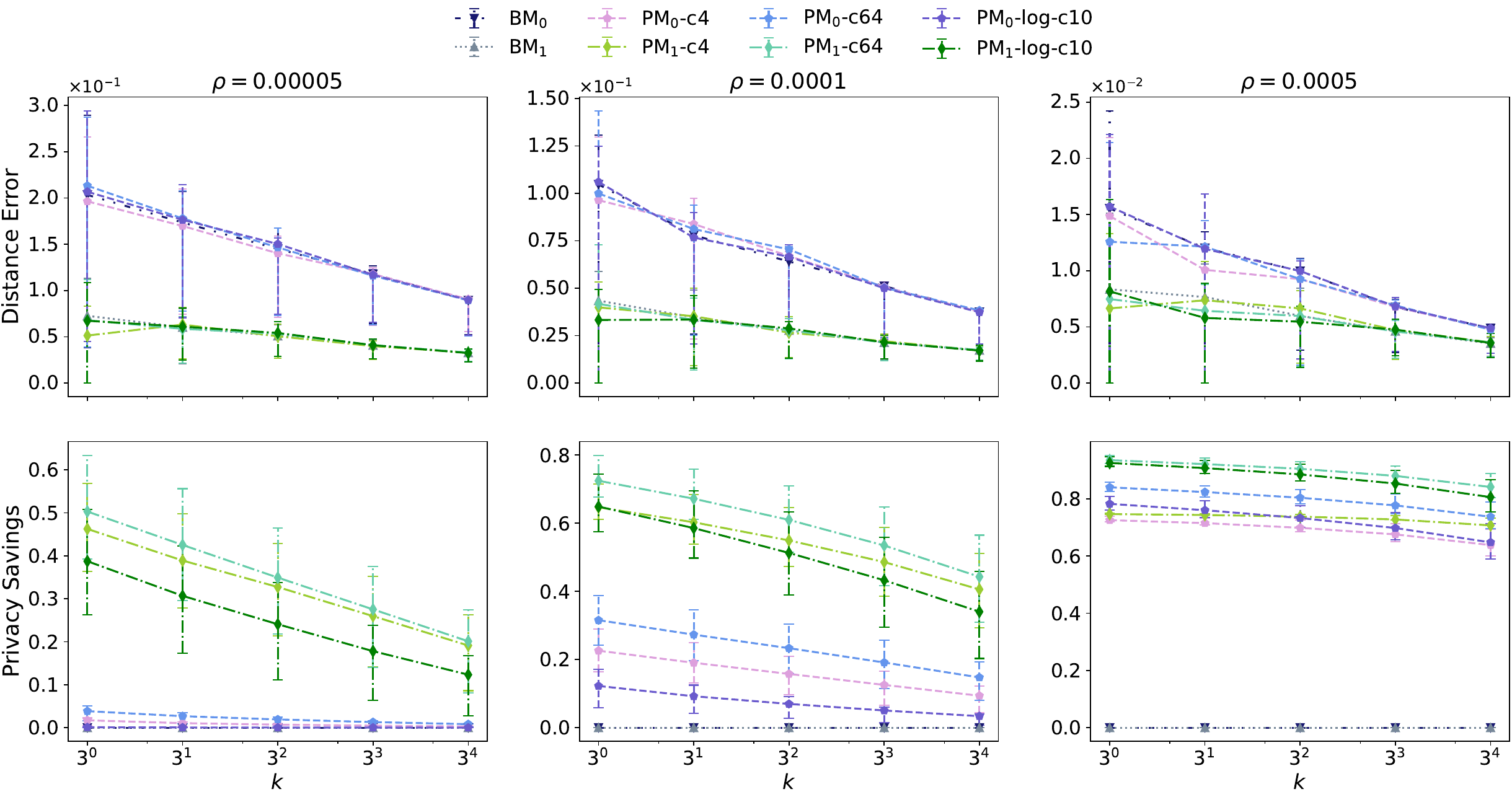}
          \caption{Single query privacy savings: $k$NN  for various $k$ and privacy parameter $\rho$.}  
        \label{fig:knn_1} 
\end{figure}
\paragraph{Single query: $k$NN} (Fig.~\ref{fig:knn_1}) 
The $k$NN experiments are performed on the MNIST dataset, which consists of images of handwritten digits represented as vectors in $[255]^d$ with $d=784=28\times 28$. 
In each experiment we use $n=60,000$ images from the training set, and find among them the $k$ nearest neighbors to an image $p_j$ uniformly drawn from the test set.
Let $J^*_j$ represent the indices of the true $k$ nearest neighbors to $p_j$, let $\tilde{J}_j$ represent a privately computed set of indices, for $j\in [300]$. For a set $S$ containing $k$ indices,  let $\mathrm{Dist}(x_S,p):=\sum_{i\in S}\|x_i-p\|$ denote the sum of the distances of the points in $x_S$ (subset of points with indices in $S$) to $p$. 
We compute the relative distance error for $k$NN:
\[
\mathrm{DistErr}(\tilde{J}_j)=\frac{\mathrm{Dist}(\tilde{J}_j,p_j)-\mathrm{Dist}(J^*_j,p_j)}{\mathrm{Dist}(J^*_j,p_j)}.
\]
We observe similar errors in the $\mathrm{PM}$ methods, compared to the respective $\mathrm{BM}$ methods. The methods using distance privatization outperform those that use point privatization, because the privatization error in $\|x_i-p\|$ in terms of distance is $O(1)$ for $\|x_i-p\|+Z_1$, while it is $O(\sqrt{d})$ for $\|x_i+Z_d-p\|$. Privacy savings in the $\mathrm{PM}$ methods depend on $k$ in this query, because less points are eliminated in each iteration for a larger $k$, as more points are closer to the smallest $k$ (noisy) distance estimates. Also, we see that using a larger $c$ (doing more rounds of eliminations) leads to more privacy savings.
\subsection{Multiple Queries utilizing Privacy Savings}
In this subsection, we demonstrate the potential utility gain over multiple queries when we utilize privacy savings. We perform multiple range queries on the T-Drive trajectory dataset, where the budgeting is done as in Algorithm~\ref{algo:mult_ie}. 
\begin{figure}[ht]
     \centering
\includegraphics[width=0.98\textwidth]{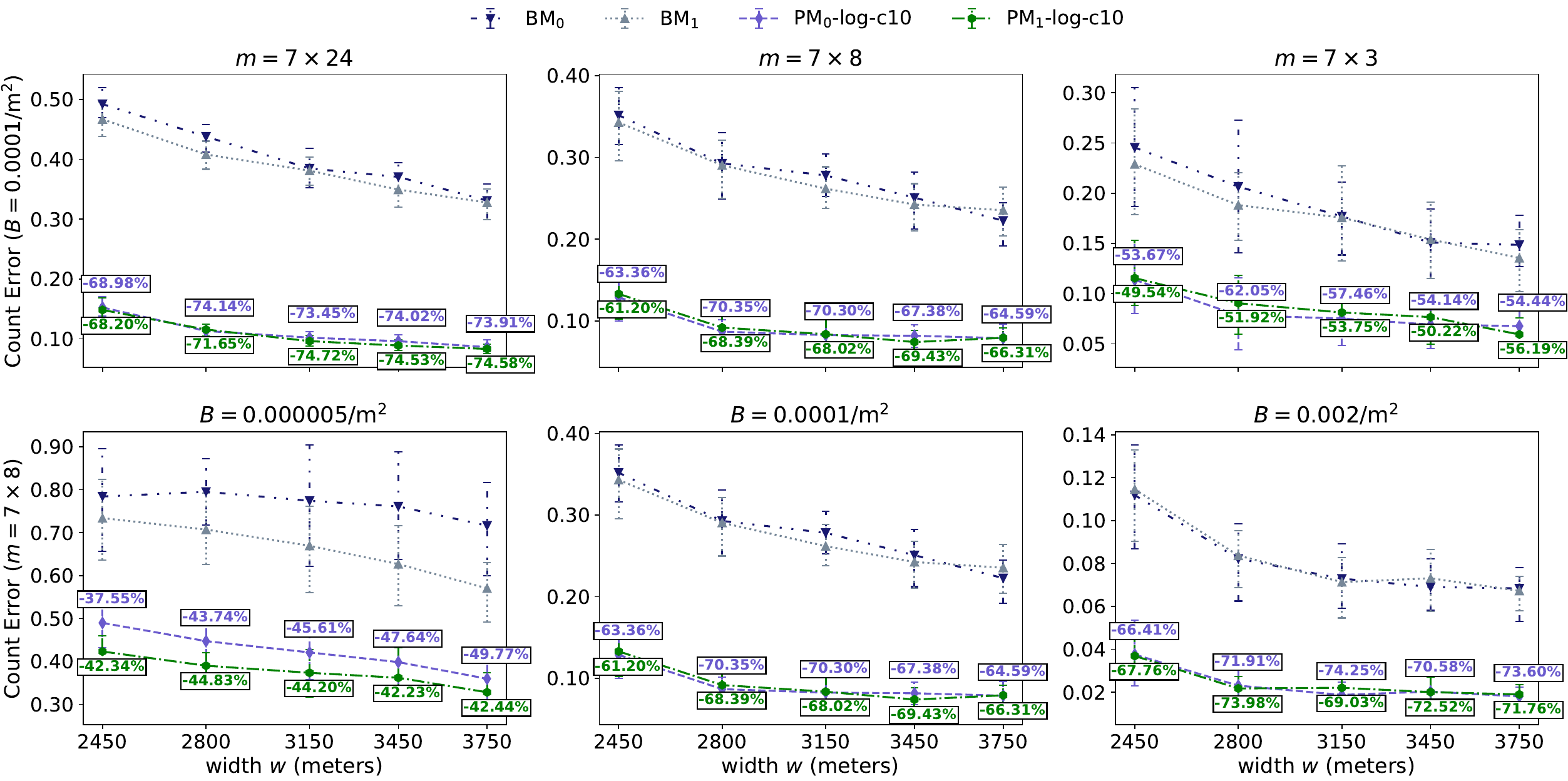}
         \caption{Multiple queries: Range counting for various range widths $w$. On the top: error w.r.t. different number of queries $m$, fixing initial budget at $B=0.0001/\mathrm{m}^2$. On the bottom: error w.r.t.
         to different initial budgets $B$, fixing $m=7\times 8$.}
        \label{fig:rc_all} 
\end{figure}
In each experiment, we make $m$ range counting queries for different ranges at $m$ time steps and calculate the average count error over the $m$ queries. The mean, the $25$-th and $75$-th percentiles reported in Fig.~\ref{fig:rc_all} are computed from repeating each experiment $50$ times, where we also label the average improvements of the $\mathrm{PM}$ methods over their baseline counterparts (\lq$-$\rq\;indicates improvement for $\mathrm{CountErr}$). We see significant utility improvements in the $\mathrm{PM}$ methods. In particular, as the number of queries $m$ increases, the improvement gap increases since more privacy savings are accumulated. Increasing the initial budget $B$ also leads to larger improvements, as we see in the single query experiments that a larger privacy parameter for one query allows for more savings, which also accumulate over multiple queries. Overall, the advantages of strategic budgeting for multiple queries are evident in the experiments.

\bibliographystyle{alpha}
\bibliography{dp}

\appendix
\section{Computing $\ProjG$}
\label{appendix:projgamma}
Let $\Gamma$ be a rectangle defined by vertices $u, v, w, z$ such that the line segments $\overline{uv}$ and $\overline{wz}$ are parallel, while $\overline{uw}$ and $\overline{vz}$ are parallel. For $S\subseteq\mathbb{R}$, write  $\lambda_S(u,v):=\{au+(1-a)v: a\in S\}$. Then $\lambda_{\mathbb{R}}(u,v)$ is the line containing $u,v$ while $\lambda_{[0,1]}(u,v)$ corresponds to the line segment $\overline{uv}$. Let $\lambda_S(u,w), \lambda_S(v,z)$ and $\lambda_S(w,z)$ be defined similarly. To determine if a point $x$ lies within $\Gamma$, we compute the projections of $x$ onto each of $\lambda_{\mathbb{R}}(u,v), \lambda_{\mathbb{R}}(u,w), \lambda_{\mathbb{R}}(v,z)$ and $\lambda_{\mathbb{R}}(w,z)$. If the projections are within the corresponding line segments, then $x$ is inside $\Gamma$. That is, we compute
\[
a_{u,v}:=\argmin_{a\in\mathbb{R}} \|au+(1-a)v-x\|
\]
for $(u,v)$ and similarly compute $a_{u,w},a_{v,z}$ and $a_{w,z}$,
and check if $0\le a_{u,v}, a_{u,w},a_{v,z},a_{w,z} \le 1$. If so, then $x$ is within $\Gamma$; otherwise, it is outside $\Gamma$.

To compute the above, note that $\argmin_{a\in\mathbb{R}} \|au+(1-a)v-x\|=\argmin_{a\in\mathbb{R}} \|au+(1-a)v-x\|^2$. 
By computing the second-order derivative w.r.t. $a$, we see that
the objective function $a\mapsto \|au+(1-a)v-x\|^2$ is convex and we can solve the minimization problem by setting the first derivative w.r.t. $a$ of the function to $0$. Doing so we get
\[
a_{u,v} = \frac{(x_1-v_1)(u_1-v_1)+(x_2-v_2)(u_2-v_2)}{(u_1-v_1)^2+(u_2-v_2)^2}.
\]
We also compute $a_{u,w}, a_{v,z}$ and $a_{w,z}$ similarly. Next, we determine the distance of $x$ to $\partial \Gamma$, which is the minimum among the distances of $x$ to each of the line segments $\overline{uv},\overline{wz},\overline{uw}$ and $\overline{vz}$.
Let $c_{u,v}$ denote the value obtained by clipping $a_{u,v}$ to $[0,1]$. Then the distance of $x$ to $\overline{uv}$ can be computed as
\[
\min_{y\in \lambda_{[0,1]}(u,v)}\|y-x\|=\|c_{u,v} u+(1-c_{u,v}) v-x\|.
\]
We can similarly compute the distances of $x$ to other line segments.

\subsection{Lipschitzness of $\ProjG$ (Lemma~\ref{lm:projgamma_lipschitz})}
\begin{proof}
Fix $u, u'\in U$. There are four cases to consider.
Let $z_u\in \partial\Gamma$ be the point such that $\|u-z_u\|=\min_{z\in \partial\Gamma} \|u-z\|$. Let $z_{u'}$ be similarly defined.

1. $u,u'\notin \Gamma$. Then $\ProjG(u)-\ProjG(u')= \|u-z_u\| - \|u'-z_{u'}\| \le \|u-z_{u'}\|-\|u'-z_{u'}\| \le \|u-u'\|$. Similarly,
$\ProjG(u')-\ProjG(u)= \|u'-z_{u'}\| - \|u-z_u\| \le \|u'-z_{u}\|-\|u-z_u\| \le \|u-u'\|$. Thus, $|\ProjG(u)-\ProjG(u')|\le 1\cdot \|u-u'\|$.

2. $u,u'\in \Gamma$. 
\begin{align*}\ProjG(u)-\ProjG(u') &= -\min_{z\in \partial\Gamma} \|u-z\| - (-\min_{z\in \partial\Gamma} \|u'-z\|)\\
&= \|u'-z_{u'}\|-\|u-z_u\|\le \|u'-u\|\end{align*} as shown in the first case. Similarly, $\ProjG(u')-\ProjG(u)= \|u-z_u\|-\|u'-z_{u'}\|\le \|u-u'\|$ and we get $|\ProjG(u)-\ProjG(u')|\le \|u-u'\|$.

3. $u\in\Gamma, u'\notin \Gamma$. Consider drawing a line from $u$ to $u'$ and let $z_p$ be the first point on the boundary that the line crosses. Then
\begin{align*}
|\ProjG(u)-\ProjG(u')| &= |-\|u-z_u\|-\|u'-z_{u'}\||\\
&=\|u-z_u\|+\|u'-z_{u'}\|\\
&\le \|u-z_p\|+\|u'-z_p\| = \|u-u'\|,
\end{align*}
where the last inequality is due to $z_p$ being on the line segment connecting $u$ and $u'$.

4. $u\notin\Gamma, u'\in \Gamma$. Following symmetric arguments to case $3$ above, $|\ProjG(u)-\ProjG(u')|=|\ProjG(u')-\ProjG(u)|\le \|u'-u\|$ 

\end{proof}

\section{Range counting: categorization of points}
\label{appendix:rc_thres_shift}
In the algorithms using point privatization, we compute $\ProjG$ on the noisy points and check if the resulting values are less than zero. Consider a point $x_i$ that is at a distance of $\gamma$ to $\partial\Gamma$. Note that if $x_i$ is inside $\Gamma$, then the noisy point $\tilde{x}_i$ is more likely to stay in $\Gamma$; if $x_i$ is outside $\Gamma$, it is also more likely that $\tilde{x}_i$ is outside $\Gamma$. To see this, draw a circle of radius slightly larger than $\gamma$ around $x_i$, check whether the circle has a larger intersection with $\Gamma$ or $\mathbb{R}^d\setminus \Gamma$.

Consider using the same method to categorize points for the algorithms using distance privatization. That is, we check whether the privatized distances are less than zero. Now, suppose the noisy distance $\tilde{y}_i$ is at a distance slightly larger than $\gamma$ to the true distance $\ProjG(x_i)$. Then it's equally likely that the point $x_i$ had been inside or outside $\Gamma$. Also, consider $A_+:=\{u\in U_l: 0<\ProjG(u)<\gamma\}$ and $A_-:=\{u\in U_l: 0>\ProjG(u)>-\gamma\}$, then the area of $A_+$ is larger than that of $A_-$; thus if points are roughly uniformly distributed around $\partial\Gamma$, then there are more points in $A_+$ than in $A_-$. In this case, if we simply check whether the privatized distances are less than zero, we will over-count the actual number of points inside $\Gamma$.
\begin{figure}[htbp]
     \centering
      \begin{subfigure}[t]{0.4\textwidth}
      \;\;\;\;\;\;\;\;\;\;\;
        \includegraphics[width=0.65\textwidth]{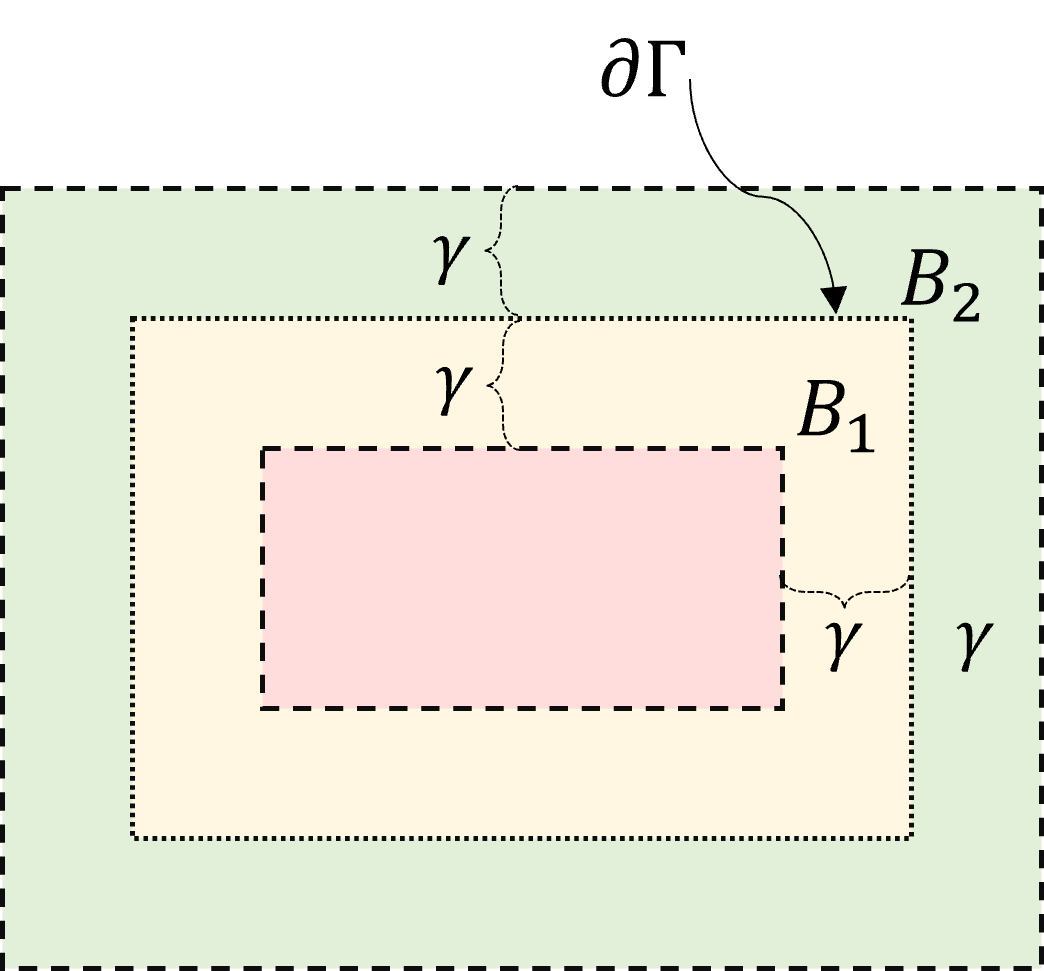}
          \subcaption{$0$ as categorization threshold.}
        \label{fig:rc_thres0_cat}
         \end{subfigure}
          \;\;\;\;\;\;\;
         \begin{subfigure}[t]{0.4\textwidth}
            \;\;\;\;
\includegraphics[width=0.65\textwidth]{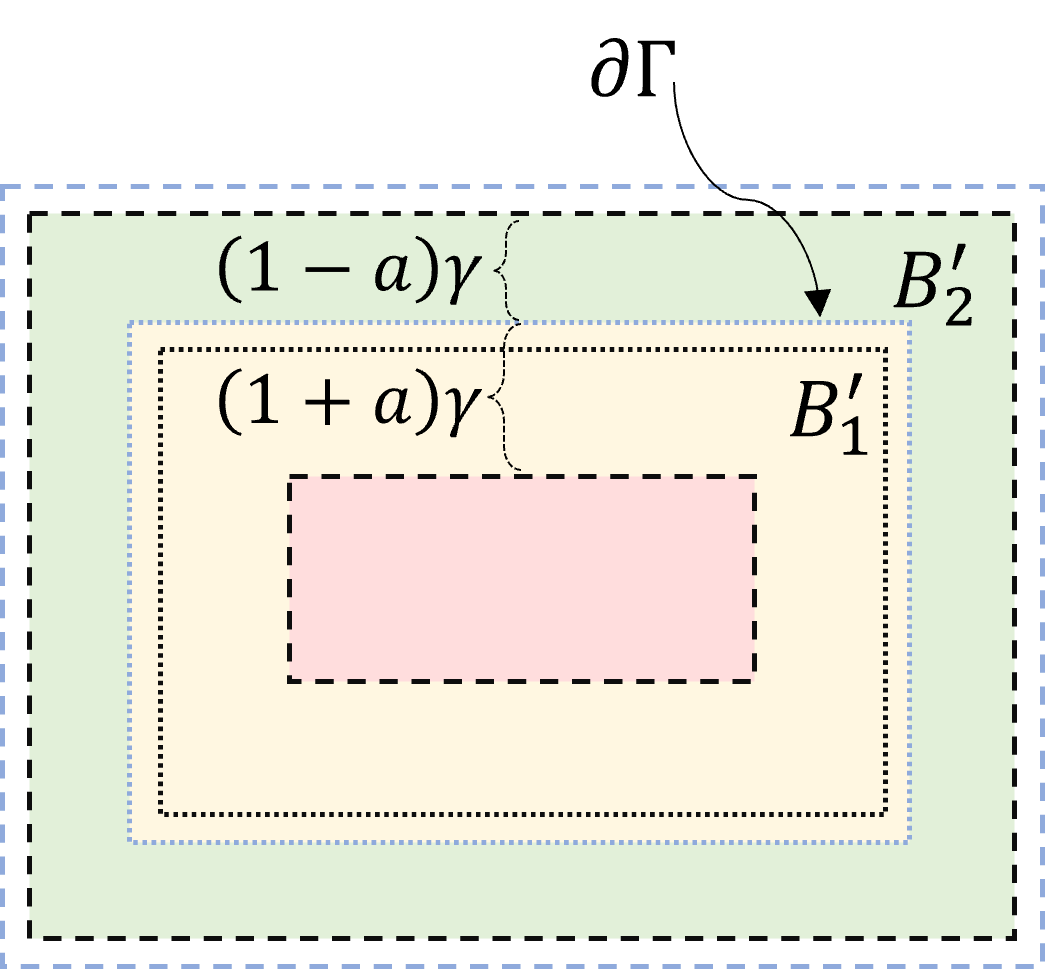}
    \subcaption{Categorization threshold shifted by $a\gamma$.} 
    \label{fig:rc_thres_eta_cat}
         \end{subfigure}
         \caption{Illustration for bands $B_1 ,B_2, B_1'$ and $B_2'$. Green: region outside $\Gamma$ where points have constant probability of being incorrectly categorized. Yellow: region inside $\Gamma$ where points have constant probability of being incorrectly categorized. On the right: after shifting, green and yellow regions have the same area. }
         \label{fig:cat_rect_shift}
\end{figure}

To mitigate this issue, we want to shrink the area of $A_+$ and expand the area of $A_-$ so that they are approximately equal. Instead of using $0$ as the categorization threshold, we will use $\eta(\gamma):=-a\gamma$. That is, we check whether $\tilde{y}_i<-a\gamma$. 
Consider the shifted regions $A_+(-a\gamma):=\{u\in U_l: -a\gamma < \ProjG(u) < (1-a)\gamma\}$ and $A_-(-a\gamma):=\{u\in U_l: -a\gamma > \ProjG(u)>-(1+a)\gamma\}$. Next, we describe a procedure to determine $a$ for rectangles. We will use intersections of rectangles to approximate $A_+(-a\gamma)
$ and $A_-(-a\gamma)$ (in reality, $A_+(-a\gamma)$ has round corners instead of square corners).

Fix $\gamma, l, w \in \mathbb{R}_{>0}$. First, we assume $2\gamma < l, w$. Consider the following three rectangles with the same center: $R$ has length $l$ and width $w$; $R_{1}$ has length $l-2\gamma$ and width $w-2\gamma$; and $R_{2}$ with length $l+2\gamma$ and width $w+2\gamma$. Consider the bands $B_2:=R_2\setminus R$ and $B_1:=R\setminus R_1$. Then $B_2$ has area $(l+2\gamma)\cdot(w+2\gamma)-l\cdot w$ and $B_1$ has area $l\cdot w -(l-2\gamma)\cdot(w-2\gamma)$. We want to shrink $B_2$ and expand $B_1$ so they have equal area (see Fig.~\ref{fig:cat_rect_shift} for an illustration); in particular we want to shift the edges of $R_2$ and $R_1$ toward the center by $a\gamma$ from all sides, which gives the new bands $B_2'$ with area 
\[(l+2(1-a)\gamma)(w+2(1-a)\gamma)-l\cdot w=2(1-a)
\gamma(l+w)+4(1-a)^2\gamma^2,\]
and $B_1'$ with area 
\[l\cdot w -(l-2(1+a)\gamma)(w-2(1+a)\gamma) = 2(1+a)\gamma (l+w) - 4(1+a)^2\gamma^2.\]
Equating the two areas above
and solving for $a$, we get one root which is in $[0,1]$, i.e. \[a=\frac{4\gamma(l+w)-4\gamma\sqrt{(l+w)^2-16\gamma^2}}{16\gamma^2}.\]

When $2\gamma\ge l$ or $2\gamma\ge w$, we set the area of $R_1$ to be zero. Then $B_1'$ has area $l\cdot w$. Proceeding as before, we solve 
\[
2(1-a)
\gamma(l+w)+4(1-a)^2\gamma^2 = l\cdot w ,
\]
obtaining $a=\frac{8\gamma^2+2\gamma(l+w)-2\gamma\sqrt{(l+w)^2+4\cdot l \cdot w}}{8\gamma^2}$. In summary, the categorization threshold is given by the function
\[
\eta(\gamma):=\begin{cases}
-\left(\frac{4\gamma(l+w)-4\gamma\sqrt{(l+w)^2-16\gamma^2}}{16\gamma}\right), \text{\;\;\;\;\;if\;} l, w > 2\gamma\\
-\left(\frac{8\gamma^2+2\gamma(l+w)-2\gamma\sqrt{(l+w)^2+4\cdot l \cdot w}}{8\gamma}\right), \text{\;\;\;else.}
\end{cases}
\]

\begin{figure}
     \centering
\includegraphics[width=0.98\textwidth]{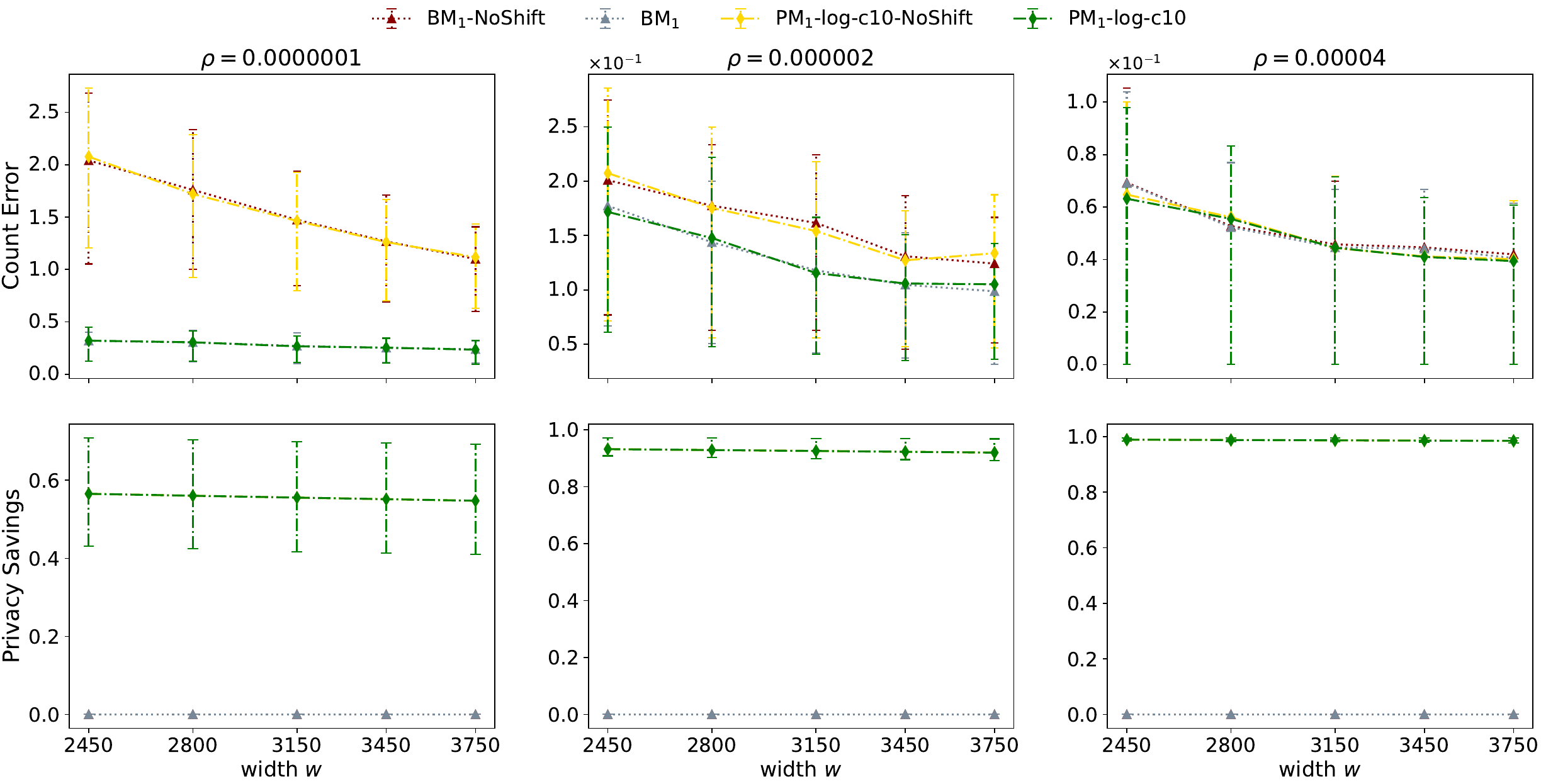}
         \caption{Comparison of distance-based methods for range counting with and without shifting the threshold (suffixed with ``NoShift"), for square ranges with various $w$ and $\rho$.}
         \label{fig:rc_comp_B}
\end{figure}

Finally, the width $\gamma$ in the discussion above is set to be equal to the scale of the Gaussian noise used in the privatization of $\ProjG(x_i)$ (i.e. for each $i$, $\eta_i=\eta(1/\sqrt{2\rho_i})$), since the noise has expected noise magnitude $O(\gamma)$ and with probability $2/3$ the noise magnitude will not be larger than $\gamma$. In Fig.~\ref{fig:rc_comp_B}, we compare the methods with and without shifting the threshold (checking the noisy distance against $\eta(\gamma)$ vs. $0$). We see that for smaller privacy budgets (where $\gamma$ is large), shifting the threshold can significantly improve performance. 

{
\section{Application in the central model}
\label{app:cdpcount}
Suppose the space $U$ is equipped with the Hamming metric. Our framework fits the classic $(\varepsilon,\delta)$-DP setting by setting $n=1$, where there is a single user corresponding to a central data curator holding some tuple $x_1$ contributed by many individuals. In the most general setting, we could have many different data curators, and an analyst is allowed to interact with any subset of them.

In this section, we discuss an application in the classic $(\varepsilon,\delta)$-DP setting (i.e. $n=1$). For cleaner composition, here we work with $\rho$-CDP, which implies $(\varepsilon,\delta)$-DP for some $\varepsilon={O}(\rho+\sqrt{\rho\log(1/\delta)})$.

\paragraph{Threshold query.} Suppose the analyst is interested in learning whether a certain fraction of households make an annual income of $\$10,000$ or less. Specifically, for a fixed $q\in (0,1)$, in a population of $N$ households, we want to compute the truth value of $f:x_1\mapsto \mathbb{1}\{|H(x_1)| < qN\}$ where $|H(x_1)|$ is the number of households in $x_1$ with income $\$10,000$ or less. A simple baseline method for this query is to first privatize the count $|H(x_1)|$, which has sensitivity $1$ (i.e. $1$-Lipschitz w.r.t. the Hamming metric), then check whether the privatized count exceeds $qN$. Let $M:(u,\rho)\mapsto |H(u)|+\frac{1}{\sqrt{2\rho}}Z$ where $Z\sim \mathcal{N}(0,1)$. We have the baseline method 
\[\mathrm{BM_1}(x_1):= \mathbb{1}\{\tilde{y} < qN\}\]
where $\tilde{y}:=M(x_1,\rho_1)\sim \mathcal{N}(|H(x_1)|,\frac{1}{2\rho_1})$.
    
For this query, the analyst does not need a very accurate estimate of $|H(x_1)|$. Ideally, he/she wants to expend as little as possible from the privacy budget. To this end, we can use the (non-interactive) iterative elimination algorithm $\mathrm{PIE}$-$\mathrm{NI}$ (see details in Algorithm~\ref{algo:threshold_pie}): We initialize $G_0=\{1\}$, let $\phi(\cdot)=|H(\cdot)|$, $\nu_{\mathrm{low}}=qN=\nu_{\mathrm{high}}$, and $r_{1,[c]}$ with $\sum_{j\in[c]} r_{1,j} = \rho_1$.
 We use the triple $(M(\cdot,\cdot;|H(\cdot)|),g_1,h_1)$ from Corollary~\ref{cor:g1h1}, where $M$ privatizes $|H(\cdot)|$ by directly adding noise to it.
\begin{algorithm}[htbp]
    \caption{Private Threshold Query via Iterative Elimination}
    \label{algo:threshold_pie}
        \begin{flushleft} 
        \textbf{Input}: 
        $x_1$; $q\in (0,1)$; number of rounds $c\in \mathbb{Z}_{> 0}$;  $\beta_0 > 0$; $r_{1,[c]} > 0$; privacy budget $B_{1}$\\
        \textbf{Output}: $ A\in \{\mathrm{Yes}$, $\mathrm{No}\}$
        \end{flushleft}
        \begin{algorithmic}[1]
        \State $N\gets |x_1|, \;\;\phi(\cdot) \gets |H(\cdot)|$
        \State $M(\cdot,\rho) \gets \phi(\cdot)+\frac{1}{\sqrt{2\rho}}Z$, \;$Z\sim\mathcal{N}(0,1)$
        \State $S_0, S_1, G_{\hat{j}}, \tilde{v}_1\gets \mathrm{PIE}\text{-}\mathrm{NI}(\{1\},c,\beta_0,r_{1,[c]},\phi, M,g_1,h_1,qN,qN, B_{1})$
        \If{$ G_{\hat{j}} == \emptyset$}
            \IfThenElse{$1 \in S_0$} {$A \gets \mathrm{Yes}$}{$A \gets \mathrm{No}$}
            % \EndIf
        \Else
            \State $\bar{y} \gets \frac{1}{\rho_1}\sum_{s\le c} r_{1,s}\tilde{v}_1(s)$ where $\rho_1=\sum_{j\in [c]} r_{1,j}$
            \State $A \gets \mathbb{1}\{\bar{y}< qN\}$
        \EndIf
        \State output $A$
        \end{algorithmic}
    \end{algorithm}
 
  Suppose we run Algorithm~\ref{algo:threshold_pie}, and get outputs $S_0,S_1,G_{\hat{j}},\tilde{v}_1$ from $\mathrm{PIE}$-$\mathrm{NI}$. Note that $S_0$, $S_1$ correspond to instances where $|H(\cdot)|$ is less than and greater than $qN$, respectively. If $G_{\hat{j}}=\emptyset$, then $S_0\cup S_1= \{1\}$. 
  By Lemma~\ref{lm:pie_non_utility}, if $1\in S_0$, then $|H(x_1)|<qN$ (with high probability); otherwise $|H(x_1)|>qN$. Thus, 
  we have:
 \begin{corollary}
\label{cor:cdpcount_utility}
    Suppose $G_{\hat{j}}=\emptyset$ in Algorithm~\ref{algo:threshold_pie}. Then, with probability $1-\beta_0$, Algorithm~\ref{algo:threshold_pie} correctly computes $f(x_1)$.
\end{corollary}

Otherwise, $\mathrm{PIE}$-$\mathrm{NI}$ completes $\hat{j}=c$ iterations, and $G_c=\{1\}$. we compute $\bar{y}:= \frac{1}{\rho_1}\sum_{s\le c} r_{1,s}\tilde{v}_1(s)$, the weighted average of all $c$ estimates of $|H(x_1)|$. Since $\bar{y}\sim \mathcal{N}(|H(x_1)|,\frac{1}{2\rho_1})$, 
We do not lose utility compared to the baseline $\mathrm{BM}_1$ in this (bad) case.

\paragraph{Experimental evaluation.} We evaluate Algorithm~\ref{algo:threshold_pie} above using household income data from the 2023 US Census dataset \cite{census2022cali}. The results reported in Fig.~\ref{fig:cdpcount1} are the mean, the $25$-th and $75$-th percentiles computed from $500$ repetitions of the same experiment. In each repetition we randomly draw $N=10000$ households from the dataset, and compute the number of households among these with income $\$10,000$ or less. We vary the value of $q$ in the threshold, and the privacy parameter $\rho$. We see that across different settings, our algorithm does not lose out in the utility compared to the baseline, while allowing us to reap significant privacy savings when the privacy parameter is sufficiently large.
}
\begin{figure}[htbp]
     \centering
\includegraphics[width=0.98\textwidth]{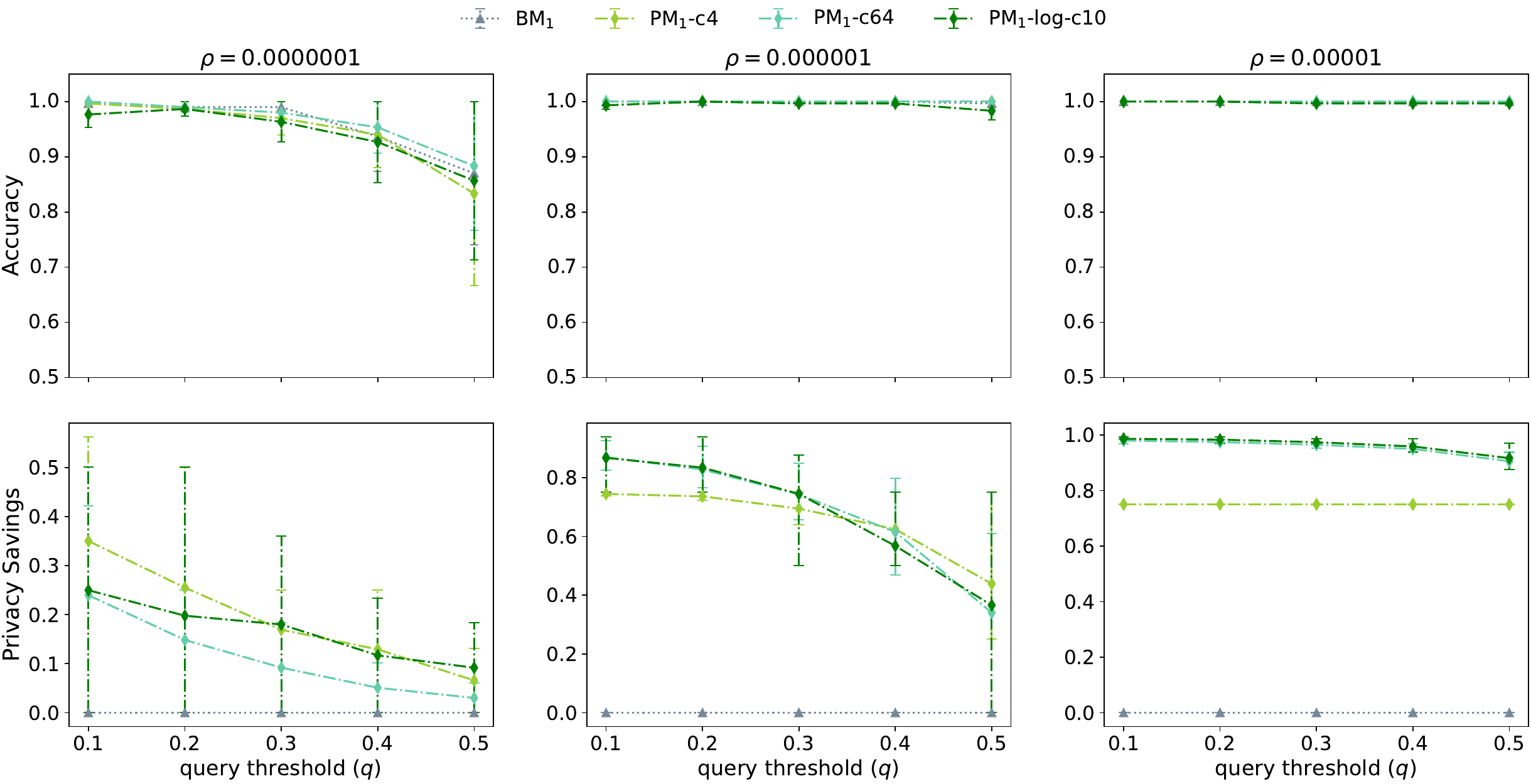}
         \caption{Single query privacy savings: threshold queries under central $\rho$-CDP for various $q$ and privacy parameter $\rho$.}
         \label{fig:cdpcount1}
\end{figure}

\section{Missing proofs}
\subsection{Proof for Theorem~\ref{thm:fulladapt_comp_GP}}
\label{appendix:fulladapt_comp_GP}
\begin{proof}
We verify the stated inequality in a straightforward manner:
\begin{align*}
\frac{m_{[t]}(x;w_{[t-1]},y_{[n],[t-1]})(z_{[t]})}{m_{[t]}(x';w_{[t-1]},y_{[n],[t-1]})(z_{[t]})} 
&= \frac{m_1(x;w_0)(z_1)\dotsb m_t(x;w_{t-1},y_{[n],[t-1]})(z_t)}{m_1(x';w_0)(z_1)\dotsb m_t(x';w_{t-1},y_{[n],[t-1]})(z_t)}\\
&=\Pi_{j=1}^t \frac{m_j(x;w_{j-1},y_{[n],[j-1]})}{m_j(x';w_{j-1},y_{[n],[j-1]})},\\
L(M_{[t]},x,x')(z_{[t]})
&=\log\left(\frac{m_{[t]}(x;w_{[t-1]},y_{[n],[t-1]})(z_{[t]})}{m_{[t]}(x';w_{[t-1]},y_{[n],[t-1]})(z_{[t]})}\right)\\
&=\sum_{j=1}^t \log\left(\frac{m_j(x;w_{j-1},y_{[n],[j-1]})(z_j)}{m_j(x';w_{j-1},y_{[n],[j-1]})(z_j)}\right)\cdot\frac{\dist(x,x')}{\dist(x,x')}\\
&\le \sum_{j=1}^t \sup_{z} \log\left(\frac{m_j(x;w_{j-1},y_{[n],[j-1]})(z)}{m_j(x';w_{j-1},y_{[n],[j-1]})(z)}\right)\cdot\frac{\dist(x,x')}{\dist(x,x')}\\
&\le \sup_{u\sim_{\Lambda}u'} \left(\frac{\sup_{z} \log\left(\frac{m_j(u;w_{j-1},y_{[n],[j-1]})(z)}{m_j(u';w_{j-1},y_{[n],[j-1]})(z)}\right)}{\dist(u,u')}\right)\dist(x,x')\\
&=\dist(x,x')\cdot\sum\nolimits_{j=1}^t \mathcal{E}_{\Lambda}({M}_j(w_{j-1},y_{[n],[j-1]})).
\end{align*}
\end{proof}

\subsection{Proof for Theorem~\ref{thm:cgp_comp_bound}}
\label{appendix:filter_cgp}
\begin{proof}
Fix a pair $x\sim_{\Lambda}x'\in U$ and an $\alpha > 1$. Let $W_{[t]}$ denote the random vector of independent coin tosses, and $Y_{[n],[t]}$ denote the collection of random vectors corresponding to outputs of all users up to time $t$. Let $Z_{[t]}$ be the random vector corresponding to the outputs of applying $M_j$ to $x$ for $j\in [t]$.
We want to compute the expectation of 
\[
H_{t}(x,x'):=e^{(\alpha-1)L\left(M_{[t]}(W_{[t-1]},Y_{[n],[t-1]}),x,x'\right)(Z_{[t]})}.
\]
Note that each $M_{j}$ is $\mathcal{F}_{j-1}$ measurable since the algorithm $M_j$ is decided at time $j-1$. Thus, for $j\le t$
\begin{align*}
&{} \frac{1}{\alpha-1}\log\left(\mathbb{E}\left[e^{(\alpha-1)L\left(M_{j}(W_{[j-1]},Y_{[n],[j-1]}),x,x'\right)(Z_{j})}|\mathcal{F}_{j-1}\right]\right) \\
&= D_{\alpha}\left(\mathcal{M}_{j}(W_{[j-1]},Y_{[n],[j-1]})(x)\|\mathcal{M}_{j}(W_{[t-1]},Y_{[n],[t-1]})(x')\right).
\end{align*}
To reduce clutter, we write $M_j=M_{j}(W_{[j-1]},Y_{[n],[j-1]})$ and omit writing $W_{[j-1]},Y_{[n],[j-1]}$ every time.
Next, we show that the ``discounted" process $\{Q_j(x,x')\}_{j\ge 0}$ defined as follows is supermartingale: Let $Q_0(x,x')=1$; for $j>0$ let $Q_j(x,x')=Q_{j-1}(x,x')\cdot D_j(x,x')$ 
where 
\begin{align*}
&D_j(x,x')=e^{(\alpha-1)L\left(M_j,x,x'\right)(Z_j)-\alpha (\alpha-1)R_j \dist(x,x')^2}.
\end{align*}
Each $D_j(x,x')$ is $F_j$-measurable. Also,
% \begin{figure*}
\begin{align*}
&\mathbb{E}[D_j(x,x')|\mathcal{F}_{j-1}] \\
&=\mathbb{E}[e^{(\alpha-1)L\left(M_j,x,x'\right)(Z_j)}|\mathcal{F}_{j-1}]\cdot e^{-\alpha(\alpha-1) R_j \dist(x,x')^2}\\
&=e^{(\alpha-1)D_{\alpha}(\mathcal{M}_j(x)\|\mathcal{M}_j(x')}\cdot e^{-\alpha(\alpha-1) R_j \dist(x,x')^2}\\
&=e^{(\alpha-1)D_{\alpha}\left(\mathcal{M}_j(x)\|\mathcal{M}_j(x')\right)\frac{\alpha\dist(x,x')^2}{\alpha\dist(x,x')^2}}\cdot e^{-\alpha(\alpha-1) R_j \dist(x,x')^2}\\
&\le e^{(\alpha-1)R_j\cdot\alpha\dist(x,x')^2}\cdot e^{-\alpha(\alpha-1) R_j \dist(x,x')^2}=1
\end{align*}
% \end{figure*}
where the first equality is due to $R_j$ being $\mathcal{F}_{j-1}$-measurable, and the inequality is due to the definition of $R_j$. 
Also, 
\begin{align*}
\mathbb{E}\left[Q_j(x,x')|\mathcal{F}_{j-1}\right] &= \mathbb{E}\left[Q_{j-1
}(x,x')\cdot D_j(x,x')|\mathcal{F}_{j-1}\right] \\
&= Q_{j-1}(x,x')\cdot \mathbb{E}[D_j(x,x')|\mathcal{F}_{j-1}]\\
&\le Q_{j-1}(x,x')
\end{align*}
for all $j>0$. Applying the inequality recursively and conditioning on $\mathcal{F}_{j-2},\mathcal{F}_{j-3},\dotsb,\mathcal{F}_{0}$, we get $\mathbb{E}\left[|Q_j(x,x')|\right]=\mathbb{E}\left[Q_j(x,x')\right]\le Q_0(x,x')=1$ for all $j>0$ since $Q_j(x,x')\ge 0$. Thus, $Q_j(x,x')$ is a supermartingale. Next,
\begin{align*}
H_{t}(x,x')&=e^{(\alpha-1)L\left(M_{[t]}(W_{[t-1]},Y_{[n],[t-1]}),x,x'\right)(Z_{[t]})} \\
&= \Pi_{j\in [t]} e^{(\alpha-1)L\left(M_j,x,x'\right)(Z_j)} \\
&=\Pi_{j\in [t]} \left( D_j(x,x')\cdot e^{\alpha(\alpha-1)R_j\dist(x,x')^2}\right)\\
&= \Pi_{j\in [t]} D_j(x,x') \cdot \Pi_{j\in [t]}e^{\alpha(\alpha-1)R_j\dist(x,x')^2}\\
&= Q_{t}(x,x')\cdot e^{\sum_{j\in [t]} \alpha(\alpha-1)R_j\dist(x,x')^2},
\end{align*}
\begin{align*}
\mathbb{E}\left[H_{t}(x,x')\right]\cdot e^{-\alpha(\alpha-1)B\dist(x,x')^2} 
&= \mathbb{E}\left[Q_{t}(x,x') e^{\underset{j\in [t]}{\sum} \alpha(\alpha-1)R_j\dist(x,x')^2}e^{-\alpha(\alpha-1)B\dist(x,x')^2}\right]\\
&= \mathbb{E}[\underbrace{Q_{t}(x,x')}_{\ge 0} \underbrace{e^{\left(\sum_{j\in [t]} R_j - B\right)\cdot \alpha(\alpha-1)\dist(x,x')^2}}_{\ge 0 \text{\;and\;} \le 1 \text{\;since\;} \sum_{j\in [t]}R_j \le B}] \\
&\le \mathbb{E}\left[Q_{t}(x,x')\right] \le 1\\
\mathbb{E}\left[H_{t}(x,x')\right] &\le e^{\alpha(\alpha-1)B\dist(x,x')^2}.
\end{align*}
Since 
\begin{align*}
\mathbb{E}[H_{t}(x,x')]=\mathbb{E}[e^{(\alpha-1)L(M_{[t]},x,x')(Z_{[t]})}]= e^{(\alpha-1)D_{\alpha}\left(\mathcal{M}_{[t]}(W_{[t-1]},Y_{[n],[t-1]})(x)|\mathcal{M}_{[t]}(W_{[t-1]},Y_{[n],[\tau-1]})(x')\right)},
\end{align*}
we get 
\begin{align*}
D_{\alpha}\left(\mathcal{M}_{[t]}(W_{[t-1]},Y_{[n],[\tau-1]})(x)|\mathcal{M}_{[t]}(W_{[t-1]},Y_{[n],[t-1]})(x')\right)\le \alpha B \dist(x,x')^2,
\end{align*}
which holds for arbitrary $x, x'$ and all $\alpha > 1$.
\end{proof}

\subsection{Proof for Theorem~\ref{thm:filter_cgp}}
\begin{proof}
    Fix a local view at user $i$ as given in Algorithm~\ref{algo:local_view_randomtau}, a pair $x\sim_{\Lambda}x'\in U$. For $t\ge 0$ let $J(t;W_{[t-1]},Y_{[n],[t-1]}):=\{j\in [t]: r_j>0\}$ denote the set of indices corresponding to the $M_j$'s that have been applied to user $i$ up to time $t$. If an $M_j$ has not been received by the local protocol, it can be identified with $M_{\mathrm{NULL}}$, where $\mathcal{R}_{\Lambda}(M_{\mathrm{NULL}})=0$ and $L(M_{\mathrm{NULL}},x,x')(\mathrm{NULL})=0$.
    Let $\tau(A)$ denote the time the analyst decides to halt the interactions in Algorithm~\ref{algo:template}. Since the decision of whether to halt at time $t$ is based on information available by time $t$, $\tau({A})$ is a stopping time. 
    Let $\tau(i)$ denote the time such that $F_B(r_{1},\dotsb,r_{\tau-1})=\mathrm{CONT}$ and $F_B(r_{1},\dotsb,R_{\tau})=\mathrm{HALT}$ at the local view. Then $\tau({i})$ is a stopping time, and so $\tau:=\min(\tau(A),\tau({i}))$ is also a stopping time. Fix any $\alpha > 1$. For $t\ge 0$, let $\{Q_j(x.x')\}_{j\ge 0}$ and $\{H_j(x.x')\}_{j\ge 0}$ be the processes as constructed in the proof of Theorem~\ref{thm:cgp_comp_bound}, where we identify the $M_j$'s not received by the local protocol with $M_{\mathrm{NULL}}$. As shown previously, $\{Q_j(x.x')\}_{j\ge 0}$ is a supermartingale.
    By Theorem~\ref{thm:stopped_proc_supmart}, for all $t\ge 0$, $\mathbb{E}[Q_{t}(x,x')]\le \mathbb{E}[Q_{\tau\wedge t}(x,x')] \le \mathbb{E}[Q_{0}(x,x')] = 1$. 
    \begin{align*}
 H_{\tau\wedge t}(x,x')&=Q_{\tau\wedge t}\cdot e^{\sum_{j\in[{\tau\wedge t}]}\alpha (\alpha-1)R_j\dist(x,x')^2}\\
 &= Q_{\tau\wedge t}\cdot e^{\sum_{j\in J(\tau\wedge t)}\alpha (\alpha-1)R_j\dist(x,x')^2}, \\
 % \end{align*}
 % \begin{align*}
    \mathbb{E}\left[H_{\tau\wedge t}(x,x')\right]\cdot e^{-\alpha(\alpha-1)B\dist(x,x')^2}
    &= \mathbb{E}\left[Q_{\tau\wedge t}(x,x')e^{\underset{j\in J(\tau\wedge t)}{\sum} \alpha(\alpha-1)R_j\dist(x,x')^2}\right]e^{-\alpha(\alpha-1)B\dist(x,x')^2}\\
&= \mathbb{E}[Q_{\tau\wedge t}(x,x')\cdot e^{\left(\sum_{j\in J(\tau\wedge t)} R_j - B\right)\cdot \alpha(\alpha-1)\dist(x,x')^2}]\\
&\le \mathbb{E}\left[Q_{\tau\wedge t}(x,x')\right] 
\le 1,\\
% \end{align*}
% \begin{align*}
\mathbb{E}\left[H_{\tau\wedge t}(x,x')\right] &\le e^{\alpha(\alpha-1)B\dist(x,x')^2}.
    \end{align*}
Let $\mathcal{P}_{[t]}(x), \mathcal{P}_{[t]}(x')$ denote the distributions of outputs from applying the $M_j$'s received at the local protocol to $x, x'$ respectively. For $t\ge \tau$, the output is $(\mathrm{HALT},\mathrm{NULL})$ so $D_{\alpha}\left(\mathcal{P}_{[t]}(x)\|\mathcal{P}_{[t]}(x')\right) = D_{\alpha}\left(\mathcal{P}_{[\tau \wedge t]}(x)\|\mathcal{P}_{[\tau \wedge t]}(x')\right)$. Also, for $t< \tau$, the output of the first coordinate is always $\mathrm{CONT}$, so does not contribute to the divergence. Thus,
\begin{alignat*}{2}
&e^{(\alpha-1)D_{\alpha}\left(\mathcal{P}_{[t]}(x)\|\mathcal{P}_{[t]}(x')\right)}
&&= e^{(\alpha-1)D_{\alpha}\left(\mathcal{P}_{[\tau \wedge t]}(x)\|\mathcal{P}_{[\tau \wedge t]}(x')\right)}\\ 
&{} &&= \mathbb{E}\left[H_{\tau\wedge t}(x,x')\right] \le e^{\alpha(\alpha-1)B\dist(x,x')^2}, \\
&D_{\alpha}\left(\mathcal{P}_{[t]}(x)\|\mathcal{P}_{[t]}(x')\right)
&&\le \alpha B \dist(x,x')^2
\end{alignat*}
which holds for arbitrary $x, x'$, all $\alpha > 1$ and all $t\ge 0$.
\end{proof}

\subsection{Proof for Theorem~\ref{thm:approxgp_filter}}
\label{appendix:approxgp_filter}
\begin{proof}
Fix a local view at user $i$. Let $J(t)$ denote the set of indices corresponding to algorithms that have been applied to $i$ up to and including time $t$. For each $t\le \tau$ where $\tau:=\min(\tau(A),\tau(i))$ and $\tau(A)$, $\tau(i)$ are as defined in the proof of Theorem~\ref{thm:filter_cgp}, we have both
\[
g_{\delta}(s_t)\sqrt{\sum_{j\in J(t)} R_j} \le B,\;\;  s_t \Lambda \sum_{j\in J(t)} R_j \le B
\]
for some $s_t>1$, so $\sum_{j\in J(t)} R_j\le B^2/g_{\delta}(s_t)^2$ and $\sum_{j\in J(t)} R_j \le B/(s_t\Lambda)$. Thus, following the arguments in the proof to Theorem~\ref{thm:filter_cgp}, 
we can show that the sequence of algorithms $M_{{J(t)}}$ satisfies $\rho:=\min\left(\frac{B^2}{g_{\delta}(s_t)^2},\frac{B}{s_t\Lambda}\right)$-CGP when restricted to pairs $x\sim_{\Lambda}x'$. If $B^2/g_{\delta}(s_t)^2\le B/(s_t\Lambda)$,
\begin{align*}
\max(g_{\delta}(s_t)\cdot \sqrt{\rho}, s_t\Lambda\rho)
=\max\left(g_{\delta}(s_t)\cdot \sqrt{B^2/g_{\delta}(s_t)^2},\; s_t\Lambda \cdot \underbrace{B^2/g_{\delta}(s_t)^2}_{\le B/(s_t\Lambda)}\right)\le B.
\end{align*}
If $B^2/g_{\delta}(s_t)^2 > B/(s_t\Lambda)$, then
\begin{align*}
\max(g_{\delta}(s_t)\cdot \sqrt{\rho}, s_t\Lambda\rho)&=\max\left(g_{\delta}(s_t)\cdot \underbrace{\sqrt{B/(s_t\Lambda)}}_{<B/g_{\delta}(s_t)},\; s_t\Lambda \cdot B/(s_t\Lambda)\right)\le B,
\end{align*}
so by Lemma~\ref{lm:cgp_to_approxgp}, $M_{J(t)}$ is $(B,\delta,\Lambda)$-GP for all $t\le \tau$.
\end{proof}

\subsection{Proof for Lemma~\ref{lm:pie_non_utility}}
\label{appendix:pie_non_utility}
\begin{proof}
Fix any $j\le c$ in the while-loop. 
Denote by $S_{0,j}, S_{1,j} \subseteq G_{j-1}$ the incremental sets of indices that get assigned to $S_0$ and $S_1$, respectively, at iteration $j$. 
For each $i\in G_{j-1}$, let $\bar{Z}_{i,j}:=\bar{\phi}_i(j)-\phi(x_i)$.
Then by the assumptions on $g(\cdot), h(\cdot)$, with probability $1-\frac{\beta_0}{c|G_{j-1}|}$, $\left|\bar{Z}_{i,j}\right|\le \bar{h}_{i,j}$. I.e., with probability $1-\frac{\beta_0}{c}$, for all $i\in G_{j-1}$ simultaneously, we have $\left|\bar{Z}_{i,j}\right|\le \bar{h}_{i,j}$.
Assuming this holds,
\begin{align*}
i\in S_{1,j} &\iff \bar{\phi}_i(j)<-\bar{h}_{i,j}+\nu_{\mathrm{low}}\\
 &\implies
\phi(x_i)=\bar{\phi}_i(j)-\bar{Z}_{i,j} \le \bar{\phi}_i(j)+\bar{h}_{i.j} < \nu_{\mathrm{low}}.
\end{align*}
Similarly, $i\in S_{0,j} \iff \bar{\phi}_i(j)>\bar{h}_{i,j}+\nu_{\mathrm{high}}$
\begin{align*}
&\implies \phi(x_i)=\bar{\phi}_i(j)-\bar{Z}_{i,j} \ge \bar{\phi}_i(j)-\bar{h}_{i,j} > \nu_{\mathrm{high}}.
\end{align*}
Taking a union bound over the $c$ iterations, $S_0=\cup_{j\le c} S_{0,j}$ and $S_1=\cup_{j\le c}S_{1,j}$ contain the correct indices after $c$ rounds with probability $1-\beta_0$.
\end{proof}

\subsection{Proof for Lemma~\ref{lm:J_subset}}
\begin{proof}
The starting subset $G_0$ contains all indices. Fix any $j\le c$ in the while-loop. For each $i\in G_{j-1}$, consider $\bar{Z}_{i,j}:=\bar{\phi}_i(j)-\phi(x_i)$.
By the definition of $\bar{h}_{i,j}$ and assumptions on $g(\cdot), h(\cdot)$, with probability $1-\frac{\beta_0}{c|G_{j-1}|}$, $\left|\bar{Z}_{i,j}\right| \le \bar{h}_{i,j}$. I.e., with probability $1-\frac{\beta_0|G_{j-1}|}{c|G_{j-1}|}=1-\frac{\beta_0}{c}$, $\left|\bar{Z}_{i,j}\right| \le \bar{h}_{i,j}$ simultaneously for all $i\in G_{j-1}$. By another union bound over the $c$ iterations in the while-loop, we have that this holds for all $j\le c$ with probability $1-\beta_0$. Next, assuming this holds, we show that $i^*_s$ is never eliminated in the while-loop (i.e. $i^*_s$ advances to $G_j$ in line $14$ for $j\le \hat{j}$), for all $s\in [k]$. Suppose for contradiction that $i^*_s$ is eliminated at some iteration $j\le \hat{j}$, for some $s\in [k]$. 
This means
\begin{equation*}
\label{eqn:is_never_elim}
    I_{i^*_s}(j)\cap I_{t_k}(j) = \emptyset \iff \bar{\phi}_{t_k}(j)+\bar{h}_{t_k,j} < \bar{\phi}_{i^*_s}(j) - \bar{h}_{i^*_s,j}. 
\end{equation*}
Also, since $i^*_s\notin \{t_1,\dotsb,t_k\}$, there is a $t_r\in G_{j-1}$ corresponding to the $r$-th smallest right end point with $r\in[k]$, such that $t_r=i^*_q$ corresponds to the $q$-th smallest $\phi(\cdot)$ value where $q>k$. Then
\begin{align*}
\bar{\phi}_{t_r}(j)+\bar{h}_{t_r,j}&\le \bar{\phi}_{t_k}(j)+\bar{h}_{t_k,j} < \bar{\phi}_{i^*_s}(j) - \bar{h}_{i^*_s,j},\\
\phi(x_{i^*_q})+\bar{Z}_{t_r,j} +\bar{h}_{t_r,j}&=\phi(x_{t_r})+\bar{Z}_{t_r,j} +\bar{h}_{t_r,j}< \phi(x_{i^*_s})+\bar{Z}_{i^*_s,j} - \bar{h}_{i^*_s,j}\\
\phi(x_{i^*_q}) &< \phi(x_{i^*_s}) - \bar{Z}_{t_r,j} - \bar{h}_{t_r,j} + \bar{Z}_{i^*_s,j}  - \bar{h}_{i^*_s,j} \\
&\le \phi(x_{i^*_s}) + \underbrace{|\bar{Z}_{t_r,j}| - \bar{h}_{t_r,j}}_{\le 0} + \underbrace{|\bar{Z}_{i^*_s,j}|  - \bar{h}_{i^*_s,j}}_{\le 0 } \le \phi(x_{i^*_s}), 
\end{align*}
which contradicts the fact that $\phi(x_{i^*_s})\le \phi(x_{i^*_q})$.
\end{proof}

\subsection{Proof for Lemma~\ref{lm:knn_BM0}}
\begin{proof} For each $i\in [n]$, let $\gamma_i:=\|\tilde{x}_i-p\|-\|x_i-p\|$. Then,
\begin{align*}
\left|\gamma_i\right|&=
\left|\|\tilde{x}_i-p\|-\|x_i-p\|\right|\le \|\tilde{x}_i-x_i\|
=  \|Z_i\|/{\sqrt{2\rho_i}},
\end{align*}
where $Z_i\sim \mathcal{N}(0,I_{d\times d})$.
So we have $|\gamma_i|\le \frac{1}{\sqrt{2\rho_i}}\lambda(d,\beta/n)$ simultaneously for $i\in [n]$ with probability $1-\beta$. We assume that this holds in the rest of the proof. Let $i^*_s$ denote the $s$-th nearest neighbor for $s\in[k]$.
First, consider $s=1$.
\begin{align*}
        \|\tilde{x}_{t_1}-p\| &\le \|\tilde{x}_{i^*_1}-p\| \\
        \|x_{t_1}-p\|+\gamma_{t_1}  &\le \|x_{i^*_1}-p\|+\gamma_{i^*_1}\\
        \|x_{t_1}-p\|- \|x_{i^*_1}-p\|&\le \gamma_{i^*_1} -  \gamma_{t_1} \le \frac{\|Z_{i^*_1}\|}{\sqrt{2\rho_{i^*_1}}}+\frac{\|Z_{t_1}\|}{\sqrt{2\rho_{t_1}}} \le \frac{\sqrt{2}}{\sqrt{\rho_0}}\cdot \lambda(d,\beta/n).
    \end{align*}
For $s\ge 2$,  we can write $t_s=\argmin_{i\in [n]\setminus \{t_1,\dotsb,t_{s-1}\}}\|\tilde{x}_i-p\|$. 
 Then, there is $q \le s$ such that $i^*_q \in [n]\setminus \{t_1,\dotsb,t_{s-1}\}$. 
    \begin{align*}
    \|\tilde{x}_{t_s}-p\| &\le \|\tilde{x}_{i^*_q}-p\|\\
    \|x_{t_s}-p\| + \gamma_{t_s} &\le \|x_{i^*_q}-p\| + \gamma_{i^*_q} \le \|x_{i^*_s}-p\| + \gamma_{i^*_q}\\
    \|x_{t_s}-p\| - \|x_{i^*_s}-p\| &\le \gamma_{i^*_q}-\gamma_{t_s} \le \frac{\|Z_{i^*_q}\|}{\sqrt{2\rho_{i^*_q}}}+\frac{\|Z_{t_s}\|}{\sqrt{2\rho_{t_s}}} \le \frac{\sqrt{2}}{\sqrt{\rho_0}}\cdot \lambda(d,\beta/n).
    \end{align*}
\end{proof}

\subsection{Proof for Lemma~\ref{lm:knn_BM1}}
\begin{proof}
    Let $\gamma_i=\frac{1}{\sqrt{2\rho_i}}Z_i$ where $Z_i\sim \mathcal{N}(0,1)$. We have with probability $1-\beta$, $|\gamma_i|\le \frac{1}{\sqrt{2\rho_i}}\lambda(1,\beta/n)$ simultaneously for $i\in [n]$. Assume this holds, then use the fact
    \[
    \|x_{t_s}-p\|+\gamma_{t_s} =\tilde{y}_{t_s} \le\tilde{y}_{i^*_s} =\|x_{i^*_s}-p\|+\gamma_{i^*_s}
    \]
    for $s\in [k]$. The rest of the proof follows the same arguments as those in the proof of Lemma~\ref{lm:knn_BM0}.
\end{proof}

\end{document}